%% file: AxiHiggs.tex
\documentclass[11pt]{article}
\usepackage[a4paper,margin=0.94in]{geometry}
\usepackage[utf8]{inputenc}
\usepackage[T1]{fontenc}
\usepackage{lmodern,amsmath,amssymb,mathtools,slashed,microtype}
\usepackage{graphicx,booktabs,array,tikz,pgfplots}
\usepackage{placeins}
\usetikzlibrary{arrows.meta,positioning,decorations.pathmorphing}
\pgfplotsset{compat=1.18}
\usepackage[colorlinks=true,linkcolor=blue,citecolor=blue,urlcolor=blue]{hyperref}
\numberwithin{equation}{section}
\newcommand{\ii}{\mathrm i}
\newcommand{\dd}{\mathrm d}
\newcommand{\vp}{v_\phi}
\newcommand{\cL}{\mathcal L}
\newcommand{\kms}{\mathrm{km\,s^{-1}}}
\newcommand{\CM}{\mathrm{CM}}
\newcommand{\lab}{\mathrm{lab}}
\newcommand{\esc}{\mathrm{esc}}
\title{\bfseries\boldmath Topological Axi-Higgs Dynamics\\
in Anomalous $U(1)$ Portals}
\author{Claudio Corian\`o$^{1,2,3}$ and Stefano Lionetti$^{1,2}$\\[0.55em]
{\small $^1$Dipartimento di Matematica e Fisica, Universit\`a del Salento}\\
{\small $^2$INFN Sezione di Lecce, Via Arnesano, 73100 Lecce, Italy}\\
{\small $^3$CNR NANOTEC, Institute of Nanotechnology, Lecce, Italy}}
\date{}
\begin{document}
\maketitle

\begin{abstract}
We develop an effective axi-Higgs model with one Standard Model
Higgs doublet, one charged hidden Higgs and an anomalous Abelian
vector portal. Extending previous Higgs--Stueckelberg constructions,
we derive the Stueckelberg kinetic term by decoupling a heavy scalar
radial mode and identify the gauge-invariant physical pseudoscalar
from the mixing of the surviving scalar phases. A single
instanton-motivated periodic interaction determines the axi-Higgs
mass, its couplings to the hidden radial excitation and the
sine-Gordon kink in the fixed-radial-field limit. We construct the
anomaly-canceling gauge completion, using an Abelian A--B example
to introduce the mechanism before its non-Abelian hidden-sector
realization. We examine the relation between the periodic potential,
the hidden gauge dynamics and the theta-vacuum energy, identifying
the invariant angle through the anomalous Yukawa phase rotation. The vector portal gives a
coherent elastic nuclear-scattering cross section and permits
additional axi-Higgs emission, whose diagrams and kinematic threshold
we determine. Numerical examples illustrate the mass and interaction
scales, and we investigate conditional thermal relic abundances and
their cosmological implications.
\end{abstract}
\newpage

\section{Introduction}
\label{sec:introduction}

Anomalous Abelian extensions of the Standard Model provide a framework
in which an axion-like particle emerges from the gauge and Higgs
structure of the theory. This particle, called the ``axi-Higgs'' in
Ref.~\cite{CIK}, arises from the mixing of a Stueckelberg field with
the phases of charged Higgs fields after spontaneous symmetry breaking.
A gauge-invariant periodic potential gives a mass to the surviving
physical CP-odd combination and determines its potential interactions.
The independent Stueckelberg parameter $M_1$ allows the mass and
couplings to depart from the familiar relations of an ordinary
Peccei--Quinn axion \cite{PQ1977,Weinberg1978,Wilczek1978}, while remaining constrained by the charges, Higgs
expectation values and anomaly coefficients of the model
\cite{CIM1,CIM2,CG2009}.
The broader connection between the axial anomaly and pseudoscalar
masses was established in Veneziano's treatment of the $U(1)$
problem in QCD~\cite{Veneziano1979}.

We realize this mass mechanism in a hidden Higgs sector whose vacuum,
spectrum and interactions follow from one explicit scalar potential. The hidden Higgs vev
determines how the physical phase is normalized. The periodic term
determines how much energy is needed to move that phase away from its
minimum. Neither ingredient alone determines the axi-Higgs mass.
We derive the scalar decoupling, interactions and scattering kinematics
for one hidden Higgs coupled to the Standard Model through a vector portal.
The phase rotation and Higgs--Stueckelberg mechanism follow earlier
work; the elementary sine-Gordon and two-body phase-space derivations
are included to make the conventions and approximations transparent.

The low-energy theory contains the Standard Model Higgs doublet $H$
and one complex hidden Higgs $\phi$, with vacuum expectation values
$v_H/\sqrt2$ and $v_\phi/\sqrt2$, respectively. We denote their
radial excitations by $h_H$ and $h_\phi$, and the dimensionless phase
of $\phi$ by $\eta/v_\phi$. The neutral phase of $H$ is absorbed by
the ordinary $Z$ boson. The explicit field parametrization is given
in Section~\ref{sec:action}.

We take the scalar potential to be the sum of a Standard Model Higgs
potential and a hidden Higgs--Stueckelberg potential. There is no
$H^\dagger H|\phi|^2$ term, and consequently no mixing of $h_H$
with $h_\phi$. Communication with ordinary matter is through the
vector portal. Mixing of $\eta$ and the Stueckelberg field produces
the physical axi-Higgs.

In the Stueckelberg formulation, the pseudoscalar $b$ shifts under the
anomalous $U(1)_B$ gauge transformation~\cite{RueggRuizAltaba2004}. Its Wess--Zumino interaction
compensates the anomalous variation of the fermionic effective action.
The effective-action framework combining Stueckelberg, axionic and
generalized Chern--Simons interactions is developed in
Refs.~\cite{ABDK2006,Anastasopoulos2007}.
When the hidden Higgs $\phi$ acquires an expectation value, its phase
$\eta$ and $b$ both enter the mass-generating sector of the gauge field
$B_\mu$. Their kinetic terms identify the combination that supplies
the longitudinal polarization of the massive vector and the orthogonal
combination that remains as the physical pseudoscalar, the axi-Higgs $\chi$. The
periodic potential then selects the constant vacuum value of $\chi$
and determines the energy required to displace it. This separates the
role of the anomaly in the gauge completion from the dynamics that
generate the particle mass.

This physical particle should be distinguished from the anomaly pole
in perturbative current correlators. Here $AVV$ denotes a correlator of
one axial and two vector currents, and $AAA$ one of three axial currents.
In the massless limit their longitudinal anomalous structures can contain
a massless pole \cite{CLM}. The Ward identity fixes its anomaly
coefficient, but does not by itself establish a propagating particle
with a specified mass. In the present construction that particle is
identified from the scalar kinetic terms, and its mass is obtained from
the potential. 

We develop this construction with $B_\mu$ as the vector portal and
$SU(2)_X$ as the hidden non-Abelian group. A hidden Dirac doublet is
vectorlike under $SU(2)_X$ but chiral under the portal. Its anomalous
gauge variation is canceled by the Stueckelberg Wess--Zumino terms.
The visible fermions carry anomaly-free $B-L$ charges. The visible
Higgs is portal neutral, so its expectation value does not enter the
rotation of the hidden phase into $\chi$.
In particular, the visible $[SU(3)_c]^2U(1)_B$ anomaly vanishes.
There is no anomaly-induced $\chi G^a_{\mu\nu}\widetilde G^{a\mu\nu}$
coupling in this construction, so it does not implement a solution
of the visible strong-CP problem.

We adapt the potential construction of Refs.~\cite{CG2009,CGLM}
to a single charged hidden Higgs. The retained operator has the
instanton-motivated form $-[\kappa\phi e^{-\ii e_Bb/M_1}+\mathrm{h.c.}]$.
Its argument is fixed by the kinetic action. Its coefficient
$\kappa$ and its linear dependence on the Higgs magnitude are
assumptions of the effective model. Defining the dimensionless
coefficient $\lambda_{\mathrm I}=\sqrt2\kappa/v_\phi^3$ gives
\begin{equation}
 m_\chi=\sqrt{\lambda_{\mathrm I}}\,v_\phi
       \sqrt{1+\frac{e_B^2v_\phi^2}{M_1^2}},
 \qquad \sigma_\chi=\frac{v_\phi M_1}{M_B}.
 \label{eq:intro-mass}
\end{equation}
This relation is conditional on the specified potential. Neither
the scalar rotation nor anomaly cancellation computes
$\lambda_{\mathrm I}$. Section~\ref{sec:hidden-dynamics} examines
this distinction quantitatively: the stated $SU(2)_X$ group remains
unbroken, and its strong dynamics prevents identifying the coefficient
with a small-size instanton weight evaluated at $v_\phi$. The mass
plots therefore scan $\lambda_{\mathrm I}$ directly. A microscopic
matching of the full confining vacuum energy is not claimed.

Changing the hidden Higgs magnitude through $h_\phi$ changes the coefficient of
$\chi^2$ in the potential and generates an $h_\phi\chi^2$ coupling.
The kinetic term also gives $h_\phi(\partial\chi)^2$. Both vertices
contribute to $h_\phi\to\chi\chi$, so their amplitudes must be
added. These interactions involve the hidden radial excitation;
the Standard Model Higgs $h_H$ remains separate.

Section~\ref{sec:ab-model} introduces the Abelian A--B model and
its anomaly cancellation. Sections~\ref{sec:action} and
\ref{sec:rotation} then establish the full field content, the action
and the physical phase. To explain the origin of $b$, we introduce a
heavy parent scalar $S$ only during the derivation. Its radial
excitation $\rho$ is removed, while its phase survives as $b$.
The fields $H$ and $\phi$ remain the two Higgs fields of the
low-energy model. We then verify the gauge transformations and
the identification of $\chi$ explicitly. Section~\ref{sec:potential} then writes the
specified periodic potential, derives the conditional mass and gives the
fixed-$h_\phi$ equation and kink. Section~\ref{sec:interactions}
derives the leading hidden interactions. The mass plots and a
numerical example are collected in Section~\ref{sec:mass-plots},
followed by the application to nuclear scattering through vector
exchange. The gauge identities needed for the portal diagrams are
given in Appendix~\ref{app:gauge-identities}.
Appendix~\ref{app:theta-vacuum} relates the Yukawa phase, the
theta angle and the vacuum energy to the cosine potential used here.

\section{The Abelian A--B model}
\label{sec:ab-model}

The Abelian A--B model of
Ref.~\cite{CIM1} contains a vector gauge field $A_\mu$, a chiral
gauge field $B_\mu$ and a Dirac fermion $\psi$. Here $A$ is the
name of a gauge field; it does not denote an axial current.
Its anomalous gauge variation is canceled by a local Stueckelberg
Wess--Zumino interaction.

We use Minkowski signature $(+---)$ and write the massless fermion
action as
\begin{equation}
 S_f[A,B]=\int\dd^4x\,\bar\psi\ii\gamma^\mu D_\mu\psi,
 \qquad
 D_\mu=\partial_\mu-\ii g_Aq_AA_\mu
 -\ii g_B(q_{BL}P_L+q_{BR}P_R)B_\mu,
 \label{eq:ab-action}
\end{equation}
where $P_L=(1-\gamma_5)/2$ and $P_R=(1+\gamma_5)/2$ project onto
left- and right-handed components. The couplings are $g_A,g_B$;
$q_A$ is the common $A$ charge of both components, and
$q_{BL},q_{BR}$ are their $B$ charges. Thus the $A$ current is
vectorlike, while the $B$ interaction is chiral when
$q_{BL}\ne q_{BR}$. The two currents coupled to the gauge fields are
\begin{equation}
 J_A^\mu=q_A\bar\psi\gamma^\mu\psi,\qquad
 J_B^\mu=\bar\psi\gamma^\mu(q_V+q_5\gamma_5)\psi,
 \qquad q_V=\frac{q_{BL}+q_{BR}}2,\quad
 q_5=\frac{q_{BR}-q_{BL}}2.
 \label{eq:ab-currents}
\end{equation}
In particular, $q_A=1$, $q_{BL}=+1$ and $q_{BR}=-1$ give a
vector $A$ current and a purely axial $B$ current, with $q_5=-1$.
The fermion transforms as
$\psi_L\mapsto e^{\ii g_Aq_A\alpha+\ii g_Bq_{BL}\omega}\psi_L$
and similarly for $\psi_R$ with $q_{BR}$, while
$A_\mu\mapsto A_\mu+\partial_\mu\alpha$ and
$B_\mu\mapsto B_\mu+\partial_\mu\omega$.

The quantum vertices are generated by the fermion determinant,
\begin{equation}
 e^{\ii\Gamma_f[A,B]}=
 \int[\dd\psi\,\dd\bar\psi]e^{\ii S_f[A,B]},
 \qquad \Gamma_f[A,B]=-\ii\,\mathrm{Tr}\log(\ii\slashed D).
 \label{eq:ab-determinant}
\end{equation}
The trace includes spacetime and spinor indices. In this expression
the gauge fields are prescribed arguments of the functional;
they remain dynamical fields in the complete action. Expanding
$\Gamma_f$ to third order in the gauge fields produces the $BAA$
triangle, a fermion loop with one $B$ and two $A$ insertions.
We choose the local counterterms so that both vector-current Ward
identities are preserved and the mixed anomaly is assigned to the
$B$ current~\cite{Bardeen:1969,ABDK2006,ArmillisCorianoGuzzi:2008}.

Let $\Gamma_{BAA}^{\lambda\mu\nu}(p,k_1,k_2)$ denote this
vertex, with all momenta incoming and $p+k_1+k_2=0$. The index
$\lambda$ belongs to $B$ and $\mu,\nu$ to the two $A$ fields.
For a massless fermion its Ward identities are
\begin{align}
 k_{1\mu}\Gamma_{BAA}^{\lambda\mu\nu}&=0,
 &k_{2\nu}\Gamma_{BAA}^{\lambda\mu\nu}&=0,\nonumber\\
 p_\lambda\Gamma_{BAA}^{\lambda\mu\nu}
 &=\mathcal A_{BAA}\epsilon^{\mu\nu\alpha\beta}
                     k_{1\alpha}k_{2\beta}.
 \label{eq:ab-ward}
\end{align}
This equation defines the tensor coefficient $\mathcal A_{BAA}$,
including charges, couplings and the overall vertex convention.
Its charge factor is $(q_{BL}-q_{BR})q_A^2$. A longitudinal
solution is
\begin{equation}
 \Gamma_{BAA,L}^{\lambda\mu\nu}
 =\mathcal A_{BAA}\frac{p^\lambda}{p^2+\ii0}
       \epsilon^{\mu\nu\alpha\beta}k_{1\alpha}k_{2\beta}.
 \label{eq:ab-pole}
\end{equation}
The remaining part of the triangle is transverse in the $B$ index.
The factor $1/p^2$ in the displayed form is the anomaly pole. It
expresses the anomalous Ward identity; a physical scalar with a
canonical kinetic term must still be identified from the scalar
action. This is the role of the Higgs--Stueckelberg analysis below.

For the explicit axial charge choice above, the left-handed anomaly
traces are
\begin{equation}
 C_{BAA}=(q_{BL}-q_{BR})q_A^2=2,\qquad
 C_{ABB}=q_A(q_{BL}^2-q_{BR}^2)=0,\qquad
 C_{BBB}=q_{BL}^3-q_{BR}^3=2.
 \label{eq:ab-traces}
\end{equation}
The pure $A$ trace vanishes. Thus both the mixed $BAA$ and the
cubic $BBB$ anomalies must be canceled. Define
$F_{A\mu\nu}=\partial_\mu A_\nu-\partial_\nu A_\mu$ and
$F_{B\mu\nu}=\partial_\mu B_\nu-\partial_\nu B_\mu$, with
$\widetilde F^{\mu\nu}=\epsilon^{\mu\nu\rho\sigma}F_{\rho\sigma}/2$
and $\epsilon^{0123}=+1$. In the prescription just specified, write
the anomalous variation as
\begin{equation}
 \delta_\omega\Gamma_f=
 \frac14\int\dd^4x\,\omega
 \left(c_{AA}F_A\widetilde F_A+c_{BB}F_B\widetilde F_B\right).
 \label{eq:ab-variation}
\end{equation}
This defines the action coefficients $c_{AA},c_{BB}$; the tensor
coefficient in Eq.~\eqref{eq:ab-ward} is obtained by differentiating
the corresponding local density, including its combinatorial
factors. Introducing a Stueckelberg field with
$b\mapsto b+M_1\omega$ gives the local completion
\begin{equation}
 \cL_{\rm kin+WZ}=-\frac14F_A^2-\frac14F_B^2
 +\frac12(\partial_\mu b-M_1B_\mu)^2
 -\frac{b}{4M_1}
       \left(c_{AA}F_A\widetilde F_A+c_{BB}F_B\widetilde F_B\right).
 \label{eq:ab-completion}
\end{equation}
The variation of the last term is exactly
$-\delta_\omega\Gamma_f$, whereas the kinetic terms are invariant.
The construction is an effective theory below its cutoff. In curved
spacetime the nonzero mixed gravitational trace also requires its
compensating term, as in the full model.

The passage to a non-Abelian vector interaction replaces $q_A$ by
a generator $T^a$, and $q_A^2$ in the mixed trace by
$\operatorname{tr}(T^aT^b)$. For one fundamental Dirac doublet of
$SU(2)_X$, this gives
$(q_{BL}-q_{BR})\operatorname{tr}(T^aT^b)=\delta^{ab}$ for
the same $B$ charges. The following section implements this step
in a model containing the Standard Model gauge group and a hidden
$SU(2)_X$ factor. The illustrative $A_\mu$ is not an extra visible
gauge boson of that model: the Standard Model representations and
the anomaly-free visible $B-L$ current are specified there. The
anomalous non-Abelian triangle relevant to the axi-Higgs is the
hidden $BXX$ vertex.

\section{The portal model and its gauge completion}
\label{sec:action}

The visible and hidden sectors communicate through the vector
$B_\mu$. It gauges a conserved visible current but is chiral on
hidden matter; its hidden anomalous variation is canceled by the
Stueckelberg Wess--Zumino interaction. The scalar potentials of $H$
and $\phi$ are separate, and their radial excitations do not mix.
We use signature $(+---)$ and $P_{L,R}=(1\mp\gamma_5)/2$.

The gauge group is
\begin{equation}
 G_{\rm SM}\times SU(2)_X\times U(1)_B,\qquad
 G_{\rm SM}=SU(3)_c\times SU(2)_L\times U(1)_Y.
 \label{eq:group}
\end{equation}
We denote the hidden-color gauge coupling by $g_X$, the portal
coupling by $g_B$, and the corresponding fields by $X_\mu^a$ and
$B_\mu$. The label $B$ denotes the portal and does not mean that
only baryon number is gauged.

To describe the scalar excitations about the vacuum, we use
electroweak unitary gauge for $H$ and polar coordinates for $\phi$:
\begin{equation}
 H=\frac{1}{\sqrt2}\begin{pmatrix}0\\v_H+h_H\end{pmatrix},
 \qquad \phi=\frac{v_\phi+h_\phi}{\sqrt2}e^{\ii\eta/v_\phi}.
 \label{eq:higgs-notation}
\end{equation}
Here $h_H$ and $h_\phi$ are the radial excitations of the two Higgs
fields, not additional Higgs multiplets. The field $\eta$ describes
the hidden phase and is canonically normalized at $h_\phi=0$.

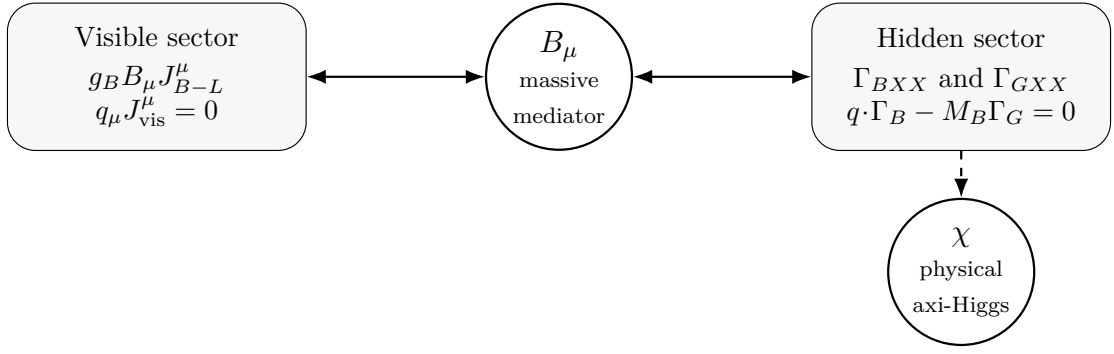
\begin{figure}[!t]
 \centering
 \resizebox{0.90\textwidth}{!}{%
 \begin{tikzpicture}[
  endcap/.style={
   draw,rounded corners=10pt,align=center,
   minimum width=3.75cm,minimum height=1.85cm,
   inner sep=6pt,fill=black!3,font=\small
  },
  bridge/.style={
   draw,circle,align=center,minimum size=1.35cm,
   inner sep=2pt,fill=white,thick
  },
  link/.style={thick,-{Latex[length=2.4mm]}}
 ]
  \node[endcap] (visible) at (0,0) {
   Visible sector\\[2pt]
   $g_BB_\mu J_{B-L}^\mu$\\
   $q_\mu J_{\rm vis}^\mu=0$
  };
  \node[bridge] (mediator) at (5.05,0) {
   $B_\mu$\\[-1pt]
   \scriptsize massive\\[-1pt]
   \scriptsize mediator
  };
  \node[endcap] (hidden) at (10.1,0) {
   Hidden sector\\[2pt]
   $\Gamma_{BXX}$ and $\Gamma_{GXX}$\\
   $q\!\cdot\!\Gamma_B-M_B\Gamma_G=0$
  };
  \node[bridge] (axion) at (10.1,-2.45) {
   $\chi$\\[-1pt]
   \scriptsize physical\\[-1pt]
   \scriptsize axi-Higgs
  };
  \draw[{Latex[length=2.4mm]}-{Latex[length=2.4mm]},thick]
   (visible) -- (mediator);
  \draw[{Latex[length=2.4mm]}-{Latex[length=2.4mm]},thick]
   (mediator) -- (hidden);
 \draw[link,dashed] (hidden) -- (axion);
 \end{tikzpicture}%
 }
 \caption{The vector connection between the two sectors. In this
 model the visible $B-L$ current is conserved, while the hidden $BXX$
 and Goldstone vertices obey the completed identity
 \eqref{eq:completed-identity}. The physical axi-Higgs $\chi$ is
 orthogonal to the eaten phase $G$. The vector is the mediator
 connecting the sectors; there is no visible--hidden Higgs coupling.}
 \label{fig:two-endcaps}
\end{figure}

Figure~\ref{fig:two-endcaps} summarizes the vector portal and the
physical phase that survives in the hidden sector. Its gauge identities
are derived in Appendix~\ref{app:gauge-identities}.

\subsection{Visible charges and neutrinos}

The visible sector is chosen so that the portal couples to the conserved
$B-L$ current. With one right-handed neutrino per generation, the charges are
\begin{equation}
 z(q_L)=z(u_R)=z(d_R)=\frac13,\qquad
 z(\ell_L)=z(e_R)=z(\nu_R)=-1,\qquad z(H)=0.
 \label{eq:visible-charges}
\end{equation}
This assignment preserves the Standard Model Yukawa interactions and
the Dirac-neutrino term $-\bar\ell_LY_\nu\widetilde H\nu_R+\mathrm{h.c.}$.
It gives the vector interaction
\begin{equation}
 \cL_{{\rm vis},B}=g_BB_\mu J_{B-L}^\mu,\qquad
 J_{B-L}^\mu=\sum_f z_f\bar f\gamma^\mu f.
 \label{eq:visible-current}
\end{equation}
The visible contributions to the local gauge and mixed gravitational
anomalies cancel. Since the visible Higgs $H$ is portal neutral, its
expectation value also leaves the portal-vector mass unchanged.

The portal charges alone allow $\phi\,\nu_R^TC\nu_R$, where $C$ is the
spinor charge-conjugation matrix. We forbid this Majorana interaction
by a separate global visible $B-L$ symmetry, under which the visible
fermions carry their usual charges and the hidden fields are neutral.
With the stated neutrino content this global symmetry is anomaly free.
It differs from the gauged portal, which also acts on the hidden fields.
This global symmetry is an imposed selection rule for the Dirac-neutrino
model, not a consequence of the portal charges; its ultraviolet
preservation is assumed.

\subsection{Hidden matter and stability}

The hidden sector contains a Dirac doublet $F$ of $SU(2)_X$, a Dirac
singlet $\Psi$, and the hidden Higgs $\phi$. Their portal charges
allow both fermion masses to arise from the hidden Higgs $\phi$:
\begin{equation}
\begin{array}{c|ccccc}
 &F_L&F_R&\Psi_L&\Psi_R&\phi\\ \hline
 SU(2)_X&\mathbf2&\mathbf2&\mathbf1&\mathbf1&\mathbf1\\
 z_B&+1&-1&+1&-1&+2\\
 D&0&0&1&1&0
\end{array}
\label{eq:hidden-charges}
\end{equation}
All three fields are Standard Model singlets. The two components of
$F$ carry hidden color, with generators $T^a=\sigma^a/2$, and its
covariant derivative is
$D_\mu=\partial_\mu-\ii g_XT^aX_\mu^a-\ii g_Bz_BB_\mu$.
The hidden-color term is absent for $\Psi$ and $\phi$. The resulting
Yukawa interactions and vacuum masses are
\begin{equation}
 \cL_Y=-y_F\phi\bar F_LF_R-y_\Psi\phi\bar\Psi_L\Psi_R
       +\mathrm{h.c.},\qquad
 m_F=\frac{y_Fv_\phi}{\sqrt2},\quad
 m_\Psi=\frac{y_\Psi v_\phi}{\sqrt2}.
 \label{eq:yukawa}
\end{equation}
Each interaction has portal charge $-1+2-1=0$. The phase of the hidden Higgs $\phi$
therefore enters both fermion masses, although only the doublet
contributes to the hidden-color anomaly.

The exact $U(1)_D$ symmetry forbids Majorana masses for
$\Psi$ and its mixing with visible neutrinos. Its mixed $D B B$
anomaly vanishes because $D_Lz_L^2-D_Rz_R^2=1-1=0$, and its
gravitational trace vanishes as well. Assuming this symmetry is
preserved by the ultraviolet completion, the lightest $D$-charged
state is stable. It can consequently serve as an incident particle
in nuclear scattering, with the incident population specified independently of the particle action.

\subsection{Scalar origin of the Stueckelberg field}
\label{sec:scalar-origin}

The Stueckelberg term can be obtained by removing the heavy radial
excitation of a charged complex scalar while retaining its phase.
This is the scalar-decoupling step used in Ref.~\cite{CG2009}.
The heavy parent field is called $S$, so that $H$ continues to denote only
the Standard Model Higgs and $\phi$ only the hidden Higgs retained
in the effective theory. The radial particle of $S$ is absent from
the low-energy spectrum; its surviving phase is the field $b$.

For definiteness, take $S$ to be a Standard Model and $SU(2)_X$
singlet with portal charge $+2$, the same charge as $\phi$. Its
covariant derivative is $D_\mu S=(\partial_\mu-\ii e_BB_\mu)S$.
The part of the parent scalar action needed for this derivation is
\begin{equation}
 \cL_S=|D_\mu S|^2-V_S(S),\qquad
 V_S=\lambda_S\left(|S|^2-\frac{w^2}{2}\right)^2,
 \qquad \lambda_S>0.
 \label{eq:parent-action}
\end{equation}
The real positive parameter $w$ fixes the magnitude of the scalar
expectation value. There is no coupling of $S$ to the Standard
Model Higgs in this construction. The purpose of this parent action
is to derive the Stueckelberg kinetic term already used below.

Near the minimum, write
\begin{equation}
 S(x)=\frac{w+\rho(x)}{\sqrt2}
                  \exp\!\left[\frac{\ii b(x)}{w}\right].
 \label{eq:parent-polar}
\end{equation}
The field $\rho$ changes the magnitude of $S$, while $b/w$
changes its phase. Both $\rho$ and $b$ have mass dimension one.
The factor $w$ in the exponent is chosen so that $b$ has a
canonical quadratic kinetic term at $\rho=0$. It does not
introduce a second normalization to be fixed later.

The radial mass is obtained directly by expanding the potential:
\begin{align}
 |S|^2-\frac{w^2}{2}&=w\rho+\frac{\rho^2}{2},\nonumber\\
 V_S&=\lambda_Sw^2\rho^2+\lambda_Sw\rho^3
                      +\frac{\lambda_S}{4}\rho^4,
 \qquad m_\rho^2=2\lambda_Sw^2.
 \label{eq:parent-radial-mass}
\end{align}
The potential has no dependence on $b$. A change of the phase at
fixed magnitude therefore produces no mass term at this stage.
This observation concerns the scalar potential; whether a phase is
a physical particle still depends on the gauge field and the other
scalar phases.

To evaluate the kinetic term, differentiate the two factors in
Eq.~\eqref{eq:parent-polar} separately:
\begin{equation}
 D_\mu S=\frac{e^{\ii b/w}}{\sqrt2}
 \left[\partial_\mu\rho+
       \ii\left(1+\frac{\rho}{w}\right)
                   (\partial_\mu b-e_BwB_\mu)\right].
 \label{eq:parent-derivative}
\end{equation}
The two terms in square brackets are respectively real and
imaginary. Their cross terms cancel when taking the absolute square,
which leaves
\begin{equation}
 |D_\mu S|^2=\frac12(\partial_\mu\rho)^2
       +\frac12\left(1+\frac{\rho}{w}\right)^2
                    (\partial_\mu b-M_1B_\mu)^2,
 \qquad M_1=e_Bw.
 \label{eq:parent-kinetic}
\end{equation}
The mass parameter appearing in the Stueckelberg term is thus the
gauge coupling times the parent scalar expectation value. In the
low-energy action it may be used as the independent parameter $M_1$;
its value is not fixed by $v_\phi$, the expectation value of the
remaining hidden Higgs.

The gauge transformation of the parent field is
$S\mapsto e^{\ii e_B\omega}S$. Equation~\eqref{eq:parent-polar}
then implies
\begin{equation}
 \rho\mapsto\rho,\qquad b\mapsto b+e_Bw\omega
                  =b+M_1\omega.
 \label{eq:parent-shift}
\end{equation}
Together with $B_\mu\mapsto B_\mu+\partial_\mu\omega$, this
gives the cancellation
\begin{equation}
 \partial_\mu(b+M_1\omega)
       -M_1(B_\mu+\partial_\mu\omega)
       =\partial_\mu b-M_1B_\mu.
 \label{eq:parent-invariance}
\end{equation}
Thus gauge invariance is already manifest before the heavy radial
mode is removed. The operation of fixing its magnitude does not
remove the shifting phase.

Consider processes with momenta and light masses much smaller than
$m_\rho$. They cannot produce the radial particle as a real final
state, and its response to the light fields is suppressed by its
large mass~\cite{AppelquistCarazzone1975}. At leading order we put $\rho=0$ in
Eq.~\eqref{eq:parent-kinetic}. The result is
\begin{equation}
 \cL_S\longrightarrow\frac12(\partial_\mu b-M_1B_\mu)^2.
 \label{eq:stueckelberg-limit}
\end{equation}
Here and below the decoupling approximation refers to the radial
mode $\rho$, not to the phase $b$. Both terms in the invariant
combination must be retained. Setting $b=0$ as well would instead
be a gauge choice and would conceal how the symmetry is realized.

To calculate the leading radial response, define
$K_\mu=\partial_\mu b-M_1B_\mu$. The terms linear and
quadratic in $\rho$ are
\begin{equation}
 \cL_S=\frac12K_\mu K^\mu
       +\frac12(\partial\rho)^2-\frac12m_\rho^2\rho^2
       +\frac{\rho}{w}K_\mu K^\mu+\cdots.
 \label{eq:rho-response-action}
\end{equation}
Consequently its linearized equation is
\begin{equation}
 (\Box+m_\rho^2)\rho=\frac{K_\mu K^\mu}{w}+\cdots,
 \qquad
 \rho\simeq\frac{K_\mu K^\mu}{w m_\rho^2}
 \quad\hbox{for slowly varying fields}.
 \label{eq:rho-response}
\end{equation}
Putting this leading solution back into the action gives
\begin{equation}
 \cL_S=\frac12K_\mu K^\mu
       +\frac{(K_\mu K^\mu)^2}{2w^2m_\rho^2}+\cdots.
 \label{eq:rho-correction}
\end{equation}
The second term is suppressed by the heavy radial mass. The
Stueckelberg action used in this paper keeps the first term.
This approximation requires both small characteristic momenta
relative to $m_\rho$ and a radial displacement small compared
with $w$. No infinite scalar coupling is needed to apply it over
a specified low-energy range.

Before $\phi$ is included, the field $b$ supplies the longitudinal
polarization of $B_\mu$. There is then no physical pseudoscalar
in the $S,B_\mu$ system after $\rho$ is removed. One can see
this by choosing $\omega=-b/M_1$. The transformed field
$B'_\mu=B_\mu-\partial_\mu b/M_1$ has the mass term
$M_1^2B'_\mu B'^{\mu}/2$. Its field strength is unchanged,
because the curl of a gradient vanishes. The three polarizations
of this massive vector account for the two transverse vector
polarizations and the original scalar phase~\cite{Higgs1964}.

The procedure is therefore to remove only the heavy radial mode
$\rho$, retaining its phase $b$ in the gauge-invariant Stueckelberg
combination, and then include the hidden Higgs $\phi$ with both its
radial excitation $h_\phi$ and its phase $\eta$. The two phases
$b,\eta$ are rotated into a combination $G$ absorbed by $B_\mu$
and an orthogonal physical pseudoscalar $\chi$, as derived in
Section~\ref{sec:rotation}. The hidden radial particle $h_\phi$
remains in the low-energy action. The Standard Model Higgs $H$
is separate and does not participate in this rotation, so $H$ and
$\phi$ remain the two retained Higgs fields.

This scalar derivation explains the Stueckelberg kinetic term and
the shift of $b$. It does not cancel the anomaly of the hidden
chiral fermions. Their quantum gauge variation still requires the
Wess--Zumino interaction, whose cancellation is displayed in the
next subsection. These are two distinct steps in constructing the
same gauge-invariant effective action.

\subsection{Stueckelberg action and anomaly cancellation}

The field $b$ is a Stueckelberg pseudoscalar and $M_1>0$ is an
independent mass parameter. With $e_B=2g_B$, the relevant terms are
\begin{align}
 \cL={}&\cL_{\rm SM+\nu_R}
 -\frac14X^a_{\mu\nu}X^{a\mu\nu}
 -\frac14F_{B\mu\nu}F_B^{\mu\nu}
 +\frac12(\partial b-M_1B)^2+|D_\mu\phi|^2
 \nonumber\\
 &+\bar F\ii\slashed D F+\bar\Psi\ii\slashed D\Psi
 +\cL_Y-V_\phi
 +\cL_{\rm WZ}+\cL_{\rm mix}.
 \label{eq:action}
\end{align}
The visible current in Eq.~\eqref{eq:visible-current} is understood
in $\cL_{\rm SM+\nu_R}$, as are the terms for the visible Higgs $H$.
The hidden potential $V_\phi(\phi,b)$ is specified in
Section~\ref{sec:potential}. For use in the anomaly terms, we define
\begin{equation}
 Q_X=\frac{g_X^2}{32\pi^2}
 X^a_{\mu\nu}\widetilde X^{a\mu\nu},\qquad
 \widetilde X^{a\mu\nu}=\frac12\epsilon^{\mu\nu\rho\sigma}
 X^a_{\rho\sigma},\qquad \epsilon^{0123}=+1.
 \label{eq:hidden-density}
\end{equation}
Here $Q_X$ is a gauge-field density, not an additional field.
Under a portal gauge transformation,
\begin{equation}
 B_\mu\mapsto B_\mu+\partial_\mu\omega,\qquad
 b\mapsto b+M_1\omega,\qquad
 \phi\mapsto e^{\ii e_B\omega}\phi.
 \label{eq:gauge}
\end{equation}
The scalar kinetic and Stueckelberg terms are separately invariant.

For example, the two changes in $\partial_\mu b-M_1B_\mu$
cancel. The derivative of the gauge phase of $\phi$ is similarly
canceled by the change of $B_\mu$ in $D_\mu\phi$. These are the
classical identities. The fermionic quantum variation below is the
additional contribution for which the Wess--Zumino interaction is
required~\cite{Fujikawa1979}.

The hidden fermions have different left- and right-handed portal
charges. Their quantum gauge variation therefore has to be canceled~\cite{Adler1969,BellJackiw1969}.
To count it, a right-handed field is represented by its left-handed
charge conjugate, whose Abelian charge has the opposite sign. This
explains the differences of left and right charges in the traces below.
The factor $1/2$ in the mixed hidden-color trace is the fundamental
generator normalization, whereas the factor two in the cubic trace
counts the two hidden-color components of $F$.

Using left-handed Weyl fields, the hidden anomaly traces are
\begin{equation}
 C_{BXX}=(1-(-1))\frac12=1,\qquad
 C_{BBB}=2(1^3-(-1)^3)+(1^3-(-1)^3)=6,
 \qquad C_{B\mathrm{grav}}=6.
 \label{eq:anomaly-traces}
\end{equation}
The two Weyl doublets $F_L,F_R^c$ avoid the global $SU(2)$
anomaly; one isolated Weyl doublet would not
\cite{Witten:1982,WangWenWitten}. There is no perturbative cubic $SU(2)$
anomaly. We work on a spin spacetime and introduce no
higher-isospin matter.

The anomalous vertices follow from the fermionic effective action,
which we denote by
\begin{equation}
 \Gamma_f[B,X,\phi]=-\ii\,\mathrm{Tr}\log\mathcal D_F
                  -\ii\,\mathrm{Tr}\log\mathcal D_\Psi,
 \label{eq:determinant}
\end{equation}
The operators $\mathcal D_j$ contain the covariant kinetic terms and
Yukawa masses, and the trace runs over spacetime, spin and internal
indices. This determinant notation retains the fermions in the
physical spectrum while generating their quantum vertices. We use a
prescription that preserves the hidden-color Ward identities and
assigns the mixed anomaly to the portal current. The associated
local anomaly-routing terms are included in $\Gamma_f$
\cite{Bardeen:1969,ABDK2006,ArmillisCorianoGuzzi:2008}.

The role of the Wess--Zumino term is to cancel a gauge variation,
rather than to remove the fermion triangle from physical amplitudes~\cite{WessZumino1971}.
Since $b$ shifts by $M_1\omega$, a term proportional to $b/M_1$
has precisely the local variation needed for this cancellation.
We state the fermion variation and its compensating term in the same
normalization so that the cancellation can be checked directly.

In this prescription, the flat-spacetime gauge variation fixes the
coefficients $c_X$ and $c_B$ through
\begin{equation}
 \delta_\omega\Gamma_f=\frac14\int\dd^4x\,\omega
 \left[c_X X^a_{\mu\nu}\widetilde X^{a\mu\nu}
       +c_B F_{B\mu\nu}\widetilde F_B^{\mu\nu}\right],
 \quad c_X=\frac{g_Bg_X^2}{4\pi^2},\quad
 c_B=\frac{g_B^3}{2\pi^2}.
 \label{eq:anomaly}
\end{equation}
The Abelian coefficient is written in the consistent-current
convention. With $\widetilde F^{\mu\nu}=\epsilon^{\mu\nu\rho\sigma}
F_{\rho\sigma}/2$ and $\epsilon^{0123}=+1$, the shift of $b$ cancels
this variation through the Wess--Zumino interaction
\begin{equation}
 \cL_{\rm WZ}=-\frac b{4M_1}
 \left[c_X X^a_{\mu\nu}\widetilde X^{a\mu\nu}
       +c_B F_{B\mu\nu}\widetilde F_B^{\mu\nu}\right].
 \label{eq:wz}
\end{equation}
The mixed gravitational anomaly requires an analogous compensating
term in curved spacetime. Together these terms define a gauge-consistent
effective action below its cutoff $\Lambda_{\rm EFT}$. The displayed
chiral spectrum is therefore understood as a Wess--Zumino effective
theory; an anomaly-free fermionic ultraviolet completion is a further
model-building question. The Stueckelberg action can be introduced
directly at this level, without deriving it by decoupling heavy partner
fermions. Its periodic potential remains a separate dynamical ingredient.
The scalar-decoupling expansion requires momenta and light masses
well below $m_\rho$; with the numerical parent parameters below,
$m_\rho\simeq35.4$ TeV. The effective theory also has a cutoff
set by its anomalous interactions and its ultraviolet completion,
which can be lower. We do not identify $M_1$ or $m_\rho$ with a
computed perturbative-unitarity bound.

The cancellation is displayed in Fig.~\ref{fig:BXX-WZ-cancellation}.
The local term fixes the gauge variation; the potential of the physical
phase is specified directly by the instanton operator in Section~\ref{sec:potential}.

\begin{figure}[!t]
 \centering
 \resizebox{0.76\textwidth}{!}{%
 \begin{tikzpicture}[
  fermion/.style={thick,-{Latex[length=1.7mm,width=1.25mm]}},
  gauge/.style={
   thick,decorate,
   decoration={snake,amplitude=1.25pt,segment length=5.2pt}
  },
  scalar/.style={thick,dashed,-{Latex[length=2mm]}},
  every node/.style={font=\small}
 ]
  % mixed hidden BXX triangle
  \coordinate (t) at (-4.7,1.25);
  \coordinate (l) at (-6.0,-0.75);
  \coordinate (r) at (-3.4,-0.75);
  \draw[fermion] (t) -- (l);
  \draw[fermion] (l) -- (r);
  \draw[fermion] (r) -- (t);
  \draw[gauge] (t) -- ++(0,1.28)
   node[above] {$B_\lambda(p)$};
  \draw[gauge] (l) -- ++(-1.15,-0.80)
   node[below left=-1pt] {$X_\mu^a(k_1)$};
  \draw[gauge] (r) -- ++(1.15,-0.80)
   node[below right=-1pt] {$X_\nu^b(k_2)$};
  \node[align=center] at (-4.7,-1.58)
   {$\Gamma_{BXX,f}^{\triangle}$};

  \node[font=\Large] at (-1.95,0.10) {$+$};

  % local b Wess--Zumino vertex
  \fill (0.65,0.05) circle (2.1pt);
  \draw[scalar] (-1.05,0.05) -- (0.61,0.05)
   node[midway,above=3pt] {$b$};
  \draw[gauge] (0.69,0.05) -- ++(1.28,1.05)
   node[above right=-1pt] {$X_\mu^a(k_1)$};
  \draw[gauge] (0.69,0.05) -- ++(1.28,-1.05)
   node[below right=-1pt] {$X_\nu^b(k_2)$};
  \node[align=center] at (0.70,-1.58)
   {$\displaystyle
     -e_B\frac{b}{M_1}Q_X$};

  \node[align=center,font=\normalsize] at (-1.85,-2.72)
   {$\displaystyle
    \delta_\omega\!\left[
      \Gamma_{f,X}
      -\!\int\!\dd^4x\,
       \frac{e_B b}{M_1}Q_X
    \right]=0$};
 \end{tikzpicture}%
 }
 \[
 \delta_\omega(\Gamma_{f,X})
 =e_B\!\int\!\dd^4x\,\omega Q_X,
 \qquad
 \delta_\omega\Gamma_{{\rm WZ},X}
 =-e_B\!\int\!\dd^4x\,\omega Q_X.
 \]
 \caption{Cancellation of the mixed hidden anomaly. The two vector
 legs are the $SU(2)_X$ fields. In the normalization of
 Eq.~\eqref{eq:hidden-density}, $c_X X\widetilde X/4=e_BQ_X$,
 so the fermionic variation is canceled by the shift of
 $-e_BbQ_X/M_1$. The notation $\Gamma_{f,X}$ includes the local
 anomaly-routing prescription defined after
 Eq.~\eqref{eq:determinant}. The figure displays the anomalous
 part of the gauge identity. For massive $F$, the vertex involving the phase of the hidden Higgs
 $\phi$ also supplies the mass-insertion term in
 Eq.~\eqref{eq:completed-identity}.}
 \label{fig:BXX-WZ-cancellation}
\end{figure}

\subsection{Kinetic mixing}

The two Abelian gauge factors also permit kinetic mixing,
\begin{equation}
 \cL_{\rm mix}=-\frac\epsilon2F_Y^{\mu\nu}F_{B\mu\nu}.
 \label{eq:kinmix}
\end{equation}
For the three visible generations, the Weyl trace
$\sum d_iY_iz_i=8$ is nonzero, so visible loops regenerate this
interaction even when $\epsilon$ vanishes at one reference scale.
The tree-level expressions below use $\epsilon(\mu_0)=0$.
A nonzero value requires simultaneous diagonalization of the neutral
kinetic and mass terms before identifying the physical couplings
\cite{Holdom:1986,BKM}. This bilinear charge trace is independent of the cubic
traces that govern anomaly cancellation. The displayed scalar
potential and spectrum are defined at tree level with the visible
and hidden Higgs potentials separate.

\section{The physical pseudoscalar}
\label{sec:rotation}

Before specifying the potential, the kinetic action already identifies
the physical phase and its normalization. Both the Stueckelberg shift
and the hidden Higgs $\phi$ contribute to the vector mass through
\begin{equation}
 \mathcal L_{b\phi}=\frac12(\partial b-M_1B)^2+|D_\mu\phi|^2,
 \qquad D_\mu\phi=(\partial_\mu-\ii e_BB_\mu)\phi.
 \label{eq:kinetic-parent}
\end{equation}
The independent parameter $M_1$ fixes the Stueckelberg contribution,
while the contribution of the hidden Higgs $\phi$ follows from the covariant derivative.
Their relative size determines which combination of phases is eaten
and which remains in the particle spectrum. This is the CP-odd Higgs--Stueckelberg mixing.

The two-doublet models of Refs.~\cite{CIM2,CG2009} contain three
neutral pseudoscalar directions: the two Higgs phases and $b$.
Two combinations are absorbed by the massive neutral vectors, leaving
one physical state. Here the neutral phase of the single visible
doublet $H$ supplies the ordinary $Z$ Goldstone, while the hidden
Higgs phase and $b$ supply one Goldstone for $B_\mu$ and one physical
axi-Higgs. A single hidden Higgs $\phi$ therefore suffices. If $\phi$
and its phase were absent, $b$ alone would be absorbed by $B_\mu$,
and this minimal field content would have no physical axi-Higgs.

We parameterize the hidden Higgs $\phi$ with vacuum magnitude
$\vp/\sqrt2$ as
\begin{equation}
 \phi=\frac{\vp+h_\phi}{\sqrt2}e^{\ii\eta/\vp},\qquad
 \mu_B=e_B\vp,\qquad M_B^2=M_1^2+\mu_B^2.
 \label{eq:polar}
\end{equation}
Here $h_\phi$ changes the magnitude of the hidden Higgs $\phi$, while
$\eta/v_\phi$ changes its dimensionless phase. The field $\eta$
is normalized canonically at $h_\phi=0$. The parameter $\mu_B$ is the
Higgs contribution to the portal-vector mass, and $M_B$ is the total
vector mass in this vacuum. The hidden Higgs vacuum parameter $\vp$ is independent of
the visible Higgs vacuum parameter $v_H$. Since $H$
is portal neutral, its neutral phase is absorbed by the ordinary
electroweak vector and does not enter the physical combination below.

In polar coordinates the kinetic term of the hidden Higgs $\phi$ separates into a radial
part and a phase part,
\begin{equation}
 |D_\mu\phi|^2=\frac12(\partial_\mu h_\phi)^2
 +\frac12\left(1+\frac{h_\phi}{v_\phi}\right)^2
       (\partial_\mu\eta-\mu_BB_\mu)^2.
 \label{eq:polar-kinetic}
\end{equation}
At $h_\phi=0$, both $\eta$ and $b$ are canonically normalized. Their gauge
shifts point along $(\mu_B,M_1)$ in the $(\eta,b)$ plane, as is also
apparent from the quadratic action
\begin{equation}
 \frac12(\partial\eta)^2+\frac12(\partial b)^2
 -B_\mu\partial^\mu(\mu_B\eta+M_1b)
 +\frac12M_B^2B_\mu B^\mu.
 \label{eq:quadratic}
\end{equation}
Only the combination $\mu_B\eta+M_1b$ appears in the term linear
in $B_\mu$. Dividing it by its length $M_B$ defines the Goldstone
field $G$ with a unit kinetic coefficient. The perpendicular combination
has no such derivative mixing with the vector and remains physical.
Because the two original kinetic coefficients are equal at $h_\phi=0$,
an ordinary orthogonal rotation is sufficient:
\begin{equation}
 G=\frac{M_1b+\mu_B\eta}{M_B},\qquad
 \chi=\frac{\mu_Bb-M_1\eta}{M_B}.
 \label{eq:rotation}
\end{equation}
The inverse rotation,
\begin{equation}
 b=\frac{M_1G+\mu_B\chi}{M_B},\qquad
 \eta=\frac{\mu_BG-M_1\chi}{M_B}.
 \label{eq:inverse}
\end{equation}
gives $G\mapsto G+M_B\omega$ and $\chi\mapsto\chi$, reducing
Eq.~\eqref{eq:quadratic} to
$\frac12(\partial\chi)^2+\frac12(\partial G-M_BB)^2$.
The vector absorbs $G$, while $\chi$ has a canonical kinetic term
and remains a physical degree of freedom. This rotation identifies
the particle but does not, by itself, give it a mass. With no periodic
interaction the potential is flat along this physical phase. The
phase-dependent potential supplies its mass and an off-diagonal
$\eta b$ entry in the CP-odd mass matrix, as shown explicitly in
Eq.~\eqref{eq:phase-matrix}. Both steps are needed: identifying the
uneaten phase and determining the energy of a displacement along it.

The same rotation determines the normalization of the invariant phase
that can enter the potential:
\begin{equation}
 \frac{\eta}{\vp}-\frac{e_Bb}{M_1}
 =-\frac{M_B}{\vp M_1}\chi
 =-\frac{\chi}{\sigma_\chi},\qquad
 \sigma_\chi=\frac{\vp M_1}{\sqrt{M_1^2+e_B^2\vp^2}}.
 \label{eq:fchi}
\end{equation}
Inserting the inverse relations makes the disappearance of the gauge
direction explicit:
\begin{align}
 \frac{\eta}{\vp}-\frac{e_Bb}{M_1}={}&\left(\frac{\mu_B}{v_\phi M_B}
              -\frac{e_B}{M_B}\right)G
 -\left(\frac{M_1}{v_\phi M_B}
              +\frac{e_B\mu_B}{M_1M_B}\right)\chi\nonumber\\
 ={}&-\frac{M_1^2+e_B^2v_\phi^2}{v_\phi M_1M_B}\chi
 =-\frac{M_B}{v_\phi M_1}\chi.
 \label{eq:fchi-derivation}
\end{align}
The coefficient of $G$ vanishes because $\mu_B=e_Bv_\phi$, and the
remaining coefficient gives $\sigma_\chi$. Thus $\sigma_\chi$ is the conversion scale between the invariant phase
and the canonically normalized physical field; it is determined by
the two kinetic terms. The strength of their periodic
energy will be specified separately. We use this normalization in the single periodic operator below,
with the notation $\sigma_\chi$ of Ref.~\cite{CGLM}.

For $M_1\gg e_B\vp$, one has $\sigma_\chi\simeq\vp$ and
$\chi\simeq-\eta$. For $M_1\ll e_B\vp$, one has
$\sigma_\chi\simeq M_1/e_B$ and $\chi\simeq b$.
The hidden Higgs vacuum parameter $v_\phi$ thus remains part of the physical
normalization even when the Stueckelberg mass is independent.
We keep $M_1$ nonzero throughout.

In unitary gauge $G=0$, the remaining kinetic term is
$\frac12(\partial\chi)^2$ at $h_\phi=0$. The next section supplies the
instanton potential and expands it about $\chi=0$.

Projecting the Wess--Zumino interaction onto the physical direction gives
\begin{equation}
 \cL_{\chi,\rm WZ}=-\frac{\mu_B}{4M_1M_B}\chi\left[
       c_X\,X^a_{\mu\nu}\widetilde X^{a\mu\nu}
       +c_B\,F_{B\mu\nu}\widetilde F_B^{\mu\nu}\right]
       \qquad(G=0).
 \label{eq:physical-wz}
\end{equation}
This is the local hidden gauge interaction of the physical field.
It must be combined with fermion triangles when calculating the
corresponding gauge-boson amplitude. The mass and scalar interactions
of $\chi$ follow instead from the potential and the kinetic terms.

The conserved visible current and the completed hidden vertices obey
the gauge identities derived in Appendix~\ref{app:gauge-identities}.
They ensure that the scattering amplitudes use the same gauge-consistent
action as the mass and decay calculations.

\subsection{Gauge transformations and the surviving particle}
\label{sec:physical-gauge}

Since the magnitude of $\phi$ is gauge invariant, its polar form gives
\begin{equation}
 h_\phi\mapsto h_\phi,\qquad
 \eta\mapsto\eta+e_Bv_\phi\omega
                  =\eta+\mu_B\omega.
 \label{eq:eta-shift}
\end{equation}
The parent scalar derivation gave $b\mapsto b+M_1\omega$.
Substituting these two shifts into Eq.~\eqref{eq:rotation},
one obtains
\begin{align}
 \delta G&=\frac{M_1(M_1\omega)+\mu_B(\mu_B\omega)}{M_B}
                   =M_B\omega,\nonumber\\
 \delta\chi&=\frac{\mu_B(M_1\omega)-M_1(\mu_B\omega)}{M_B}=0.
 \label{eq:explicit-phase-shifts}
\end{align}
The two contributions to $\delta\chi$ cancel. This is the
basic reason that $\chi$ survives a gauge choice: it is unchanged
by the gauge transformation that removes $G$.

In the plane of canonically normalized fields $(\eta,b)$,
a gauge transformation changes the fields in the direction
$(\mu_B,M_1)$. The physical direction $(-M_1,\mu_B)$ is
perpendicular to it, since
\begin{equation}
 (\mu_B,M_1)\cdot(-M_1,\mu_B)=0.
 \label{eq:phase-orthogonality}
\end{equation}
Both vectors have length $M_B$. Dividing by this length gives
the normalized combinations in Eq.~\eqref{eq:rotation}.
Their orthogonality is measured with the ordinary Euclidean scalar
product in field space because the two quadratic kinetic
coefficients are both one. This field-space statement is separate
from the spacetime signature of the action.

To pass to unitary gauge choose $\omega=-G/M_B$.
The transformed fields satisfy
\begin{equation}
 G'=0,\qquad \chi'=\chi,\qquad
 B'_\mu=B_\mu-\frac{\partial_\mu G}{M_B}.
 \label{eq:unitary-choice}
\end{equation}
Since $F_B(B')=F_B(B)$, the quadratic vector and phase action
can be written as
\begin{equation}
 -\frac14F_{B\mu\nu}(B')F_B^{\mu\nu}(B')
 +\frac12M_B^2B'_\mu B'^{\mu}
 +\frac12(\partial\chi)^2.
 \label{eq:unitary-quadratic}
\end{equation}
The first two terms describe a massive vector and the last term
describes the physical pseudoscalar. Gauge fixing has not removed
$\chi$, and it has not assigned a mass to it. That mass follows
from the instanton potential in the next section.

The alternative gauge choice $b=0$ gives the same physical field. In this gauge, denote the remaining phase and vector
by $\eta$ and $B_\mu$. The phase-vector quadratic terms are
\begin{equation}
 \frac12(\partial\eta)^2
 -\mu_BB_\mu\partial^\mu\eta
 +\frac12M_B^2B_\mu B^\mu.
 \label{eq:bzero-action}
\end{equation}
The middle term shows that $\eta$ is not yet separated from
the longitudinal vector. Completing the square gives
\begin{equation}
 \frac12M_B^2\left(B_\mu-
          \frac{\mu_B}{M_B^2}\partial_\mu\eta\right)^2
 +\frac12\frac{M_1^2}{M_B^2}(\partial\eta)^2.
 \label{eq:bzero-square}
\end{equation}
The vector shift again leaves its field strength unchanged.
The remaining scalar is canonically normalized by
\begin{equation}
 \chi=-\frac{M_1}{M_B}\eta\qquad(b=0),
 \label{eq:bzero-chi}
\end{equation}
which is precisely Eq.~\eqref{eq:rotation} evaluated in this
gauge. Setting $b=0$ therefore leaves the same physical particle,
represented entirely by the remaining Higgs phase. It does not
permit a second independent gauge choice that also sets $\eta=0$.

Before removing $\rho$, the two charged complex scalars $S,\phi$ have
four real components and the Abelian gauge field has two transverse
polarizations. After symmetry breaking they are reorganized into
the three polarizations of $B_\mu$, the two radial fields
$\rho,h_\phi$, and the physical pseudoscalar $\chi$: six
degrees of freedom on either side. Removing the heavy radial
particle leaves five. The Standard Model Higgs and electroweak
sector are spectators in this count.

Gauge invariance of the interaction is checked with the same two
phase shifts. The combination used in the potential obeys
\begin{equation}
 \frac{\eta+\mu_B\omega}{v_\phi}
 -\frac{e_B(b+M_1\omega)}{M_1}
 =\frac{\eta}{v_\phi}-\frac{e_Bb}{M_1},
 \label{eq:invariant-phase-check}
\end{equation}
because $\mu_B/v_\phi=e_B$. It is independent of $G$ and
equals $-\chi/\sigma_\chi$. A potential built from this
combination can give a mass to $\chi$ while preserving the
portal symmetry. In the gauge $b=0$, its argument is simply
$\eta/v_\phi=-\chi/\sigma_\chi$, so the same potential and
mass follow in either description.

\section{The periodic potential and the axi-Higgs mass}
\label{sec:potential}

We now specify the potential directly, following the
Higgs--Stueckelberg construction of Refs.~\cite{CG2009,CGLM}.
The scalar sector contains the Standard Model Higgs $H$ and the
hidden Higgs $\phi$ defined in Eq.~\eqref{eq:higgs-notation}.
Their potentials are separate:
\begin{align}
 V(H,\phi,b)&=V_H(H)+V_\phi(\phi,b),\nonumber\\
 V_H&=-\mu_H^2 H^\dagger H+\lambda_H(H^\dagger H)^2,\nonumber\\
 V_\phi&=-\mu_\phi^2|\phi|^2+\lambda_\phi|\phi|^4+V',
 \qquad
 V'=-\left[\kappa\phi e^{-\ii e_Bb/M_1}+\mathrm{h.c.}\right].
 \label{eq:potential-full}
\end{align}
The product $\phi e^{-\ii e_Bb/M_1}$ is invariant under
Eq.~\eqref{eq:gauge}. The two factors acquire opposite phases,
so this operator gives a periodic interaction without breaking the
portal gauge symmetry. It is the one-hidden-Higgs counterpart of
the Higgs--Stueckelberg operators used in the two-doublet models
of Refs.~\cite{CG2009,CGLM}. We retain only this leading operator.

After choosing the phase origin at the angular minimum, we take
$\kappa$ real and positive and define
\begin{equation}
 \kappa=\frac{\lambda_{\mathrm I}\vp^3}{\sqrt2},\qquad
 \lambda_{\mathrm I}>0.
 \label{eq:instanton-coefficient}
\end{equation}
The parameter $\kappa$ has mass dimension three and is held fixed
when differentiating the potential. Its normalization by $v_\phi^3$
is a convention. In a controlled semiclassical matching one could
write $\lambda_{\mathrm I}=\mathcal C_{\mathrm I}e^{-S_{\mathrm I}}$,
with $S_{\mathrm I}=8\pi^2/g_X^2(\mu)$ and a prefactor containing
the determinant, size integral and Yukawa factors. No order-one
value of that prefactor follows from this notation. We do not use
this identification to infer the coefficient for the unbroken
hidden gauge group.

Removing the hidden fermion mass phase by a chiral rotation transfers that phase
to the theta term; together with the Wess--Zumino term, this gives
the invariant angle $\theta_X-\chi/\sigma_\chi$.
Appendix~\ref{app:theta-vacuum} derives this statement and the
leading instanton harmonic. Here the phase origin is chosen at the
minimum, so $\chi$ denotes the fluctuation about it and the
coefficient $\kappa$ is real and positive.

The gauge-invariant phase factor can also be obtained using the parent
scalar of Section~\ref{sec:scalar-origin}. Since $S$ and $\phi$
have the same portal charge, $S^\dagger\phi$ is gauge invariant.
At fixed parent magnitude,
\begin{equation}
 \frac{\sqrt2}{w}S^\dagger\phi
 \;\longrightarrow\;\phi e^{-\ii b/w}
       =\phi e^{-\ii e_Bb/M_1}.
 \label{eq:parent-operator}
\end{equation}
Thus the operator in Eq.~\eqref{eq:potential-full} has a simple
scalar representation before radial decoupling. A term
$-[\sqrt2\kappa S^\dagger\phi/w+\mathrm{h.c.}]$ gives the
retained interaction after the heavy radial mode is removed.
The coefficient of this scalar operator is restricted by the
relative-phase symmetry described below. Radial decoupling explains
its field dependence but does not calculate its nonperturbative
coefficient. Small changes in the heavy scalar minimum can be absorbed
into the matched value of $w$, with its response suppressed in the
decoupling expansion.

\subsection{Perturbative protection and hidden gauge dynamics}
\label{sec:hidden-dynamics}

Two distinct assumptions enter the use of a small periodic coefficient:
the absence of perturbative phase-breaking operators and the size of
the nonperturbative contribution. Gauge invariance alone fixes neither.
In the parent scalar description, $S^\dagger\phi$ is allowed and
has dimension two. An unrestricted coefficient multiplying it would
produce the same cosine after radial decoupling, without instanton
suppression.

We therefore impose a relative-phase symmetry on the perturbative
scalar and Yukawa action. One convenient choice is
\begin{equation}
 S\mapsto e^{\ii\alpha}S,\qquad
 \phi\mapsto\phi,\qquad F\mapsto F,\qquad\Psi\mapsto\Psi.
 \label{eq:relative-phase-symmetry}
\end{equation}
The displayed kinetic terms, magnitude-dependent scalar potentials
and Yukawa interactions respect this transformation. It forbids
$S^\dagger\phi$ and its nonzero powers in the perturbative potential.
Phase-independent terms such as $|S|^2|\phi|^2$ are allowed and can
be included in the matched radial parameters. In the effective action
the Wess--Zumino interaction shifts by topological densities under
Eq.~\eqref{eq:relative-phase-symmetry}; this is the anomalous breaking
that permits a nonperturbative potential. Thus the absence of a
perturbative cosine is a symmetry assumption, rather than a consequence
of the portal charges. Its preservation by an ultraviolet completion
is also an assumption. It protects against perturbative scalar
contributions, but does not ensure a small hidden strong-interaction
contribution.

This is a hidden-sector Peccei--Quinn-type symmetry: a global phase
symmetry, spontaneously broken by the scalar expectation values and
anomalously broken in the gauge action~\cite{PQ1977}. After the portal
vector eats $G$, its surviving physical direction is $\chi$.
Indeed, Eq.~\eqref{eq:relative-phase-symmetry} induces
$\chi\mapsto\chi+\sigma_\chi\alpha$, so the unit-anomaly phase
normalization is $f_\chi=\sigma_\chi$. The axi-Higgs therefore has
the interpretation of a hidden axion in this realization. The unit
hidden-color anomaly agrees with $N_{\rm DW}=1$ obtained from the
compact phase and its first harmonic below. Although the illustrative
physical field is mostly $\eta$, namely
$\chi\simeq0.119\,b-0.993\,\eta$, its shift comes from
$\delta b=w\alpha$: the exact coefficient of $b$ times
$w=25$ TeV is $\sigma_\chi\simeq2.979$ TeV. Thus the small
$b$ component is consistent with the full global phase shift. Protecting
this symmetry against additional ultraviolet breaking is the associated
axion-quality assumption. This interpretation does not couple it to
QCD or solve the visible strong-CP problem.

Both $S$ and $\phi$ are $SU(2)_X$ singlets, so their expectation
values do not break that group. Above the mass of its single Dirac
fundamental the one-loop running is
\begin{equation}
 \mu\frac{\mathrm d g_X}{\mathrm d\mu}
 =-\frac{b_0g_X^3}{16\pi^2},\qquad
 b_0=\frac{20}{3}\quad(\mu>m_F),\qquad
 b_0=\frac{22}{3}\quad(\mu<m_F).
 \label{eq:hidden-running}
\end{equation}
The second coefficient describes pure Yang--Mills theory after the
fermion threshold. Matching the coupling at $m_F$ gives, at this
order~\cite{PDGQCD2025},
\begin{equation}
 \Lambda_X=m_F\exp\!\left[-\frac{3}{22}
 \left(\frac{8\pi^2}{g_X^2(v_\phi)}
                  +\frac{20}{3}\ln\frac{m_F}{v_\phi}\right)\right].
 \label{eq:hidden-scale}
\end{equation}
For example, $v_\phi=3$ TeV, $m_F=600$ GeV and
$8\pi^2/g_X^2(v_\phi)=32$ give $\alpha_X(v_\phi)=0.196$
and $\Lambda_X\simeq33$ GeV. This is a one-loop indicator of the
strong scale, not a precise determination of a hadron mass.

There is no hidden-color Higgs expectation value to suppress large
instantons. The small-size one-instanton measure, including one
light-mass insertion, behaves schematically as
$\mathrm d\rho\,\rho^{-5}(\rho\Lambda)^{20/3}(m_F\rho)$
up to logarithms, where $\rho$ is the instanton size. It grows toward
large sizes until the approximation ceases to apply. Beyond the
fermion threshold the pure-gauge running still does not supply a
perturbative infrared cutoff. Consequently evaluating an exponential
at $\mu=v_\phi$ does not determine the full vacuum energy.

In the regime $m_F\gg\Lambda_X$, dimensional analysis instead
suggests an angular energy of order $\Lambda_X^4$. With
$\sigma_\chi\simeq2.979$ TeV its characteristic mass scale would
be of order $\Lambda_X^2/\sigma_\chi$, approximately $0.37$ GeV
for the preceding example. The numerical coefficient and angular
shape require nonperturbative input. Even the radial dependence is
different from a single mass insertion: at fixed high-scale coupling,
threshold matching gives
\begin{equation}
 \Lambda_X(r)^{22/3}=m_F(r)^{2/3}\Lambda_{1}^{20/3},
 \qquad \Lambda_X(r)^4\propto r^{4/11},
 \qquad m_F(r)=y_Fr/\sqrt2.
 \label{eq:heavy-threshold-radial}
\end{equation}
Here $\Lambda_1$ is the one-loop scale of the theory with one
active Dirac flavor. This leading heavy-fermion estimate does not
give the linear dependence on $r$ assumed in
Eq.~\eqref{eq:periodic-potential}.

We calculate the consequences of a specified, leading periodic operator; we do
not claim to have matched its amplitude or radial dependence to the
confining $SU(2)_X$ theory. In particular, the light-mass illustration
in Section~\ref{sec:mass-plots} is not a prediction for the above
choice of $g_X$. It does not exclude such masses at a smaller hidden
strong scale: Section~\ref{sec:mass-plots} gives the corresponding
coupling estimate with its normalization assumption stated explicitly.
A confining completion would have to include its
own angular energy, hidden glueballs and hidden hadrons containing
$F$. Calling the coefficient an input does not eliminate those
contributions. A quantitative matching to such a completion remains
outside the results established here.

\subsection{Mass and fixed-radial-field dynamics}

The mass and interaction calculations retain the linear radial
dependence of the specified operator. Substituting the physical
phase gives
\begin{align}
 \phi e^{-\ii e_Bb/M_1}
 &=\frac{v_\phi+h_\phi}{\sqrt2}
           e^{\ii(\eta/v_\phi-e_Bb/M_1)}
 =\frac{v_\phi+h_\phi}{\sqrt2}e^{-\ii\chi/\sigma_\chi},
 \nonumber\\
 \kappa\phi e^{-\ii e_Bb/M_1}+\mathrm{h.c.}
 &=\sqrt2\kappa(v_\phi+h_\phi)
                       \cos(\chi/\sigma_\chi).
 \label{eq:operator-to-cosine}
\end{align}
The Hermitian conjugate supplies the opposite phase. For real
$\kappa$ their sum is twice the cosine, which fixes both the
normalization and sign of the periodic term.

Using the polar form of $\phi$ and the normalization in
Eq.~\eqref{eq:fchi}, the periodic term becomes
\begin{equation}
 V'(h_\phi,\chi)=-\lambda_{\mathrm I}\vp^3(\vp+h_\phi)
                       \cos\!\left(\frac{\chi}{\sigma_\chi}\right).
 \label{eq:periodic-potential}
\end{equation}
For the chosen positive coefficient we expand about $\chi=0$.
Stationarity with respect to the two radial fields at
$h_H=h_\phi=0$ gives
\begin{equation}
 \mu_H^2=\lambda_Hv_H^2,
 \qquad \mu_\phi^2=(\lambda_\phi-\lambda_{\mathrm I})\vp^2.
 \label{eq:stationarity}
\end{equation}
Their masses follow from the second derivatives:
\begin{equation}
 m_{h_H}^2=2\lambda_Hv_H^2,
 \qquad m_{h_\phi}^2=(2\lambda_\phi+\lambda_{\mathrm I})\vp^2,
 \qquad
 \left.\frac{\partial^2 V}{\partial h_H\partial h_\phi}\right|_0=0.
 \label{eq:radial-masses}
\end{equation}
There is no radial mixing to diagonalize. In particular, $h_H$ is
the Standard Model Higgs particle and $h_\phi$ is the hidden radial
particle. Positive $\lambda_H$ and $\lambda_\phi$ keep the
quartic potential bounded below. With $\lambda_{\mathrm I}>0$,
the radial masses in Eq.~\eqref{eq:radial-masses} are positive.

In terms of the two radial fields the potential is
\begin{align}
 V={}&-\frac{\mu_H^2}{2}(v_H+h_H)^2
       +\frac{\lambda_H}{4}(v_H+h_H)^4
       -\frac{\mu_\phi^2}{2}(v_\phi+h_\phi)^2
       +\frac{\lambda_\phi}{4}(v_\phi+h_\phi)^4
       \nonumber\\
 &-\sqrt2\kappa(v_\phi+h_\phi)
                              \cos(\chi/\sigma_\chi).
 \label{eq:radial-written-out}
\end{align}
The first derivatives at $h_H=h_\phi=\chi=0$ are
\begin{align}
 \left.\frac{\partial V}{\partial h_H}\right|_0
       &=-\mu_H^2v_H+\lambda_Hv_H^3,\nonumber\\
 \left.\frac{\partial V}{\partial h_\phi}\right|_0
       &=-\mu_\phi^2v_\phi+\lambda_\phi v_\phi^3
                                     -\sqrt2\kappa.
 \label{eq:radial-first-derivatives}
\end{align}
Setting each to zero gives Eq.~\eqref{eq:stationarity}. The
hidden equation includes the linear contribution from the
instanton term. One must use this equation before replacing the
quadratic parameters by the expectation values in the masses.
For example,
\begin{equation}
 \left.\frac{\partial^2V}{\partial h_\phi^2}\right|_0
 =-\mu_\phi^2+3\lambda_\phi v_\phi^2
 =(2\lambda_\phi+\lambda_{\mathrm I})v_\phi^2.
 \label{eq:radial-second-derivative}
\end{equation}
Although the instanton term is linear in $h_\phi$, it enters
this mass through the stationary value of $\mu_\phi^2$.
There is no derivative involving both $h_H$ and $h_\phi$,
because the potential is a sum of separate visible and hidden
parts. The expectation values and masses of the two sectors can
therefore be specified without a mixing angle.

To obtain the dynamics of $\chi$ alone, we hold $h_\phi=0$ and
set the other field fluctuations to zero. Subtracting the constant
value of the potential at $\chi=0$ gives
\begin{equation}
 V_\chi(\chi)=\lambda_{\mathrm I}\vp^4
        \left[1-\cos\!\left(\frac{\chi}{\sigma_\chi}\right)\right].
 \label{eq:cosine-potential}
\end{equation}
The quadratic term is $m_\chi^2\chi^2/2$, with
\begin{equation}
 m_\chi^2=\frac{\lambda_{\mathrm I}\vp^4}{\sigma_\chi^2}
 =\lambda_{\mathrm I}\vp^2
            \left(1+\frac{e_B^2\vp^2}{M_1^2}\right).
 \label{eq:mass}
\end{equation}
Equivalently, before the rotation in Eq.~\eqref{eq:rotation},
the CP-odd mass matrix in the basis $(\eta,b)$ is
\begin{equation}
 \mathcal M_{\rm odd}^2=\lambda_{\mathrm I}\vp^4
 \begin{pmatrix}
  \vp^{-2}&-e_B/(\vp M_1)\\
  -e_B/(\vp M_1)&e_B^2/M_1^2
 \end{pmatrix}.
 \label{eq:phase-matrix}
\end{equation}
It has a zero eigenvalue along the eaten direction $G$ and the
eigenvalue $m_\chi^2$ along $\chi$. Thus the potential and the
kinetic rotation identify the same physical field.

The zero eigenvalue is a direct consequence of gauge invariance,
and can be checked without solving a quadratic characteristic
equation. Expanding the original periodic operator to second order
in the two phases gives
\begin{equation}
 V'_{\rm quadratic}
 =\frac{\lambda_{\mathrm I}v_\phi^4}{2}
       \left(\frac{\eta}{v_\phi}-\frac{e_Bb}{M_1}\right)^2.
 \label{eq:phase-quadratic-square}
\end{equation}
Differentiating this square with respect to $\eta$ and $b$
gives the entries of Eq.~\eqref{eq:phase-matrix}. The matrix
annihilates the gauge direction:
\begin{equation}
 \mathcal M_{\rm odd}^2
       \begin{pmatrix}\mu_B\\M_1\end{pmatrix}=0.
 \label{eq:mass-gauge-direction}
\end{equation}
For example, its first component contains
$\mu_B/v_\phi^2-e_B/v_\phi=0$. The second component
vanishes for the same reason, $\mu_B=e_Bv_\phi$. Moving
the phases in this direction is a gauge transformation and cannot
change their potential energy.

On the orthogonal direction the same multiplication gives
\begin{equation}
 \mathcal M_{\rm odd}^2
       \begin{pmatrix}-M_1\\\mu_B\end{pmatrix}
 =m_\chi^2\begin{pmatrix}-M_1\\\mu_B\end{pmatrix},
 \qquad
 m_\chi^2=\lambda_{\mathrm I}v_\phi^4
           \left(\frac1{v_\phi^2}+\frac{e_B^2}{M_1^2}\right).
 \label{eq:mass-physical-direction}
\end{equation}
Equivalently, inserting the inverse rotation into
Eq.~\eqref{eq:phase-quadratic-square} gives
$V'_{\rm quadratic}=m_\chi^2\chi^2/2$, with no term in
$G$. The kinetic and potential calculations therefore select
the same two combinations. The absence of a potential mass for
$G$ does not imply a second physical massless particle: $G$
is the combination absorbed by the massive vector.

The components of the physical particle are fixed by the two
contributions to the vector mass. For the parameters of
Section~\ref{sec:mass-plots}, $\mu_B=e_Bv_\phi=1.20$ TeV
and $M_1=10$ TeV, so that
\begin{equation}
 G\simeq0.993\,b+0.119\,\eta,
 \qquad \chi\simeq0.119\,b-0.993\,\eta.
 \label{eq:phase-example}
\end{equation}
In this example the eaten combination is mainly the parent
scalar phase $b$, while the physical particle is mainly the phase
of the retained hidden Higgs $\phi$. Both coefficients are
needed for the exact cancellation of the gauge shift. The
decomposition changes when $M_1/(e_Bv_\phi)$ changes, whereas
the gauge invariance of $\chi$ and its canonical normalization
remain exact. The specified potential coefficient controls the mass
of this already identified physical combination.

In terms of the dimensionless potential coefficient, the mass is
\begin{equation}
 m_\chi=\sqrt{\lambda_{\mathrm I}}\,\vp
               \sqrt{1+\frac{e_B^2\vp^2}{M_1^2}}.
 \label{eq:instanton-mass}
\end{equation}
The square root multiplying $\sqrt{\lambda_{\mathrm I}}v_\phi$
comes from canonical normalization. A small specified coefficient
gives a light field; a small coefficient is not inferred from the
gauge charges or from the phase rotation.

In the fixed-$h_\phi$ approximation the Lagrangian is simply
\begin{equation}
 \cL_\chi=\frac12\partial_\mu\chi\partial^\mu\chi-V_\chi(\chi).
 \label{eq:chi-action}
\end{equation}
Varying this action yields
\begin{equation}
 \Box\chi+m_\chi^2\sigma_\chi
                    \sin\!\left(\frac{\chi}{\sigma_\chi}\right)=0,
 \qquad \Box=\partial_t^2-\boldsymbol\nabla^2.
 \label{eq:chi-equation}
\end{equation}
For small oscillations, $|\chi|\ll\sigma_\chi$, it reduces to
$(\Box+m_\chi^2)\chi=0$. Holding the radial field fixed is an
approximation to the coupled equations, applicable when the hidden
radial response can be neglected. The hierarchy
$m_{h_\phi}\gg m_\chi$ supports it for the slowly varying
configurations considered here.

\begin{figure}[tbp]
 \centering
 \includegraphics[width=0.96\textwidth]{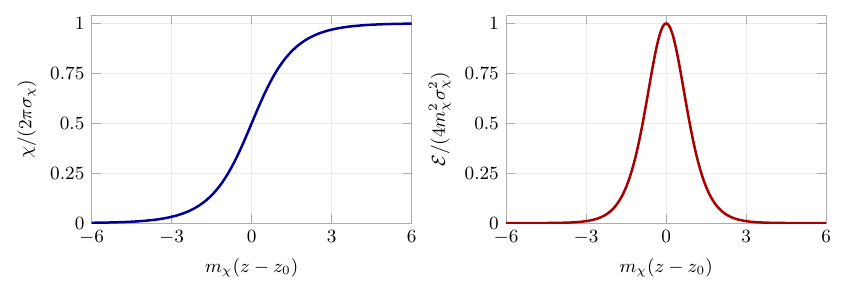}
 \caption{The kink in the fixed-$h_\phi=0$ approximation.
 Left: the physical field interpolates between neighbouring minima
 of the cosine potential. Right: its energy density is localized
 near $z_0$. Dimensionless axes separate the shape from the width
 $m_\chi^{-1}$ and field scale $\sigma_\chi$.}
 \label{fig:kink}
\end{figure}

For a static configuration depending only on $z$, the same equation
admits the kink~\cite{MantonSutcliffe2004}
\begin{equation}
 \chi_{\rm kink}(z)=4\sigma_\chi
              \arctan\!\left[e^{m_\chi(z-z_0)}\right].
 \label{eq:kink}
\end{equation}
The field interpolates from $0$ to $2\pi\sigma_\chi$ over a
distance of order $m_\chi^{-1}$. The arbitrary constant $z_0$
locates its centre. Its energy density is
\begin{equation}
 \mathcal E(z)=\frac12\left(\frac{\mathrm d\chi}{\mathrm dz}\right)^2
                 +V_\chi(\chi)
 =4m_\chi^2\sigma_\chi^2\,
                  \operatorname{sech}^2[m_\chi(z-z_0)].
 \label{eq:kink-energy}
\end{equation}
Figure~\ref{fig:kink} displays the field and the localized energy.
Both follow from the single cosine potential in
Eq.~\eqref{eq:cosine-potential} with $h_\phi$ fixed.

Write $\alpha_S=b/w$ and $\alpha_\phi=\eta/v_\phi$,
each with period $2\pi$. A portal transformation shifts both
by the same amount, so the surviving angle is
$\Theta=\alpha_\phi-\alpha_S$, with period $2\pi$.
The invariant $S^\dagger\phi$ contains the first harmonic of
this angle. There is therefore one inequivalent minimum per physical
period, $N_{\rm DW}=1$, in this compact realization. The endpoints
of the kink are the same vacuum expressed in adjacent periods.
The planar solution is meaningful in the fixed-radius theory, but
it does not establish a stable cosmological network separating
inequivalent vacua. Strings and radial excursions matter for its
formation and decay~\cite{Dine2023}. A constant $\theta_X$ translates
the minima; it does not bias their energies. We make no cosmological
abundance claim for these configurations.

For a variation $\delta\chi$ that vanishes at the boundary, integration by parts gives
\begin{equation}
 \delta S_\chi
 =\int\dd^4x\left[\partial_\mu\chi\partial^\mu\delta\chi
                         -V_\chi'(\chi)\delta\chi\right]
 =-\int\dd^4x\,[\Box\chi+V_\chi'(\chi)]\delta\chi.
 \label{eq:chi-variation}
\end{equation}
In this equation the prime means differentiation with respect to
$\chi$, and
\begin{equation}
 V_\chi'(\chi)=\frac{\lambda_{\mathrm I}v_\phi^4}{\sigma_\chi}
                   \sin(\chi/\sigma_\chi)
 =m_\chi^2\sigma_\chi\sin(\chi/\sigma_\chi).
 \label{eq:chi-force}
\end{equation}
The arbitrary variation gives Eq.~\eqref{eq:chi-equation}.
Expanding the sine to first order supplies the Klein--Gordon
equation. For example, a small plane wave has
$\omega_{\mathbf k}^2=|\mathbf k|^2+m_\chi^2$, making the
meaning of the mass extracted from the potential explicit.

The kink can be derived by one further integration. For a static
field depending on $z$, Eq.~\eqref{eq:chi-equation} becomes
\begin{equation}
 \frac{\mathrm d^2\chi}{\mathrm dz^2}
                  =\frac{\mathrm d V_\chi}{\mathrm d\chi}.
 \label{eq:static-chi-equation}
\end{equation}
Multiply by $\mathrm d\chi/\mathrm dz$ and integrate with
respect to $z$. A configuration approaching a minimum with zero
gradient at either end satisfies
\begin{equation}
 \frac12\left(\frac{\mathrm d\chi}{\mathrm dz}\right)^2
                   =V_\chi(\chi)
 =2m_\chi^2\sigma_\chi^2
                       \sin^2\!\left(\frac{\chi}{2\sigma_\chi}\right).
 \label{eq:kink-first-integral}
\end{equation}
For the increasing profile from $0$ to $2\pi\sigma_\chi$,
choose the positive square root. With
$u=\chi/(2\sigma_\chi)$, the resulting first-order equation is
\begin{equation}
 \frac{\mathrm d u}{\mathrm dz}=m_\chi\sin u,
 \qquad
 \int\frac{\mathrm du}{\sin u}
       =\ln\tan\!\left(\frac u2\right)=m_\chi(z-z_0).
 \label{eq:kink-integration}
\end{equation}
Solving for $\chi$ gives Eq.~\eqref{eq:kink}. The same
first integral shows that the gradient and potential contributions
to its energy are equal at every point. Substituting the profile
then gives the energy density in Eq.~\eqref{eq:kink-energy}.
Its integral is
\begin{equation}
 \int_{-\infty}^{+\infty}\dd z\,\mathcal E(z)
                         =8m_\chi\sigma_\chi^2.
 \label{eq:kink-tension}
\end{equation}
In three spatial dimensions this is the energy per unit transverse
area of the planar profile. This derivation uses precisely the
fixed-$h_\phi$ action of Eq.~\eqref{eq:chi-action}; it introduces
no additional potential or scalar field.

\section{Interactions of the hidden Higgs and the axi-Higgs}
\label{sec:interactions}

Only the hidden radial excitation $h_\phi$ enters the periodic term.
The Standard Model excitation $h_H$ has its usual interactions and
does not acquire an $h_H\chi\chi$ vertex from this potential.
We first keep the dependence on $h_\phi$ in
Eq.~\eqref{eq:periodic-potential}, obtaining
\begin{equation}
 V'=-\lambda_{\mathrm I}\vp^3(\vp+h_\phi)
       +\frac12m_\chi^2\chi^2
       +\frac{m_\chi^2}{2\vp}h_\phi\chi^2
       -\frac{m_\chi^2}{24\sigma_\chi^2}\chi^4+\cdots.
 \label{eq:potential-expansion}
\end{equation}
The linear radial term is already included in the stationary
condition in Eq.~\eqref{eq:stationarity}. The cubic potential
coupling is $m_\chi^2/\vp$. For the retained linear radial
operator ($p=1$ in the comparison below), there is no
$h_\phi^2\chi^2$ term in the potential. This absence is specific
to its radial dependence, not a general property of a periodic potential.

The derivative interactions are fixed by the hidden Higgs kinetic
term. In unitary gauge, substituting Eq.~\eqref{eq:inverse} into
Eq.~\eqref{eq:polar-kinetic} gives the exact expression
\begin{align}
 \cL_{b\phi}={}&\frac12(\partial h_\phi)^2
       +\frac12(\partial\chi)^2+\frac12M_B^2B_\mu B^\mu
       \nonumber\\
 &+\left(\frac{h_\phi}{\vp}+\frac{h_\phi^2}{2\vp^2}\right)
       \left(\mu_BB_\mu+\frac{M_1}{M_B}\partial_\mu\chi\right)
       \left(\mu_BB^\mu+\frac{M_1}{M_B}\partial^\mu\chi\right).
 \label{eq:physical-kinetic}
\end{align}
This expression supplies the $h_\phi\chi\chi$, $h_\phi B\chi$
and $h_\phi BB$ kinetic vertices and their terms quadratic in
$h_\phi$. In particular,
\begin{align}
 \cL_{h_\phi\chi\chi}
 &=-\frac{m_\chi^2}{2\vp}h_\phi\chi^2
       +\frac{M_1^2}{\vp M_B^2}h_\phi(\partial\chi)^2,
       \label{eq:hphichichi}\\
 \cL_{h_\phi B\chi}
 &=\frac{2\mu_BM_1}{\vp M_B}
                  h_\phi B_\mu\partial^\mu\chi.
       \label{eq:hphiBchi}
\end{align}
The potential and derivative terms in Eq.~\eqref{eq:hphichichi}
contribute to the same decay. Their on-shell coefficient is
\begin{equation}
 \mathcal G_{h_\phi\chi\chi}
 =\frac1{\vp}\left[m_\chi^2+
       \frac{M_1^2}{M_B^2}(m_{h_\phi}^2-2m_\chi^2)\right].
 \label{eq:hphi-amplitude}
\end{equation}
For $m_{h_\phi}>2m_\chi$, the partial width is therefore
\begin{equation}
 \Gamma(h_\phi\to\chi\chi)
 =\frac{|\mathcal G_{h_\phi\chi\chi}|^2}{32\pi m_{h_\phi}}
                \sqrt{1-\frac{4m_\chi^2}{m_{h_\phi}^2}}.
 \label{eq:hphi-width}
\end{equation}
The dependence on the radial assumption can be isolated without
changing the kinetic action. For comparison, let the periodic term
be $V'_p=-A r^p\cos(\chi/\sigma_\chi)$, where
$r=v_\phi+h_\phi$ and $A v_\phi^p=m_\chi^2\sigma_\chi^2$.
Its on-shell radial vertex is
\begin{equation}
 \mathcal G^{(p)}_{h_\phi\chi\chi}
 =\frac1{v_\phi}\left[p\,m_\chi^2+
 \frac{M_1^2}{M_B^2}(m_{h_\phi}^2-2m_\chi^2)\right].
 \label{eq:radial-power-vertex}
\end{equation}
The potential contains
$p(p-1)m_\chi^2h_\phi^2\chi^2/(4v_\phi^2)$, which vanishes
for $p=1$ but not for $p=4/11$. For a consistent comparison the
stationary point and the physical radial mass must be held fixed:
$\mu_\phi^2=\lambda_\phi v_\phi^2-Ap v_\phi^{p-2}$ and
$m_{h_\phi}^2=2\lambda_\phi v_\phi^2+Ap(2-p)v_\phi^{p-2}$.
For either illustrative light mass, choosing $p=1$ or $p=4/11$
gives $\Gamma(h_\phi\to\chi\chi)\simeq3.63$ GeV at the same
physical radial mass, because the derivative contribution dominates.
This local comparison does not determine the full confining potential;
the remaining explicit expansion retains $p=1$.

Expanding the square in
Eq.~\eqref{eq:physical-kinetic} gives
\begin{equation}
 \left(\mu_BB_\mu+\frac{M_1}{M_B}\partial_\mu\chi\right)^2
 =\mu_B^2B_\mu B^\mu
 +\frac{2\mu_BM_1}{M_B}B_\mu\partial^\mu\chi
 +\frac{M_1^2}{M_B^2}(\partial\chi)^2.
 \label{eq:kinetic-square-expanded}
\end{equation}
Multiplication by $h_\phi/v_\phi$ gives the three cubic
interactions. Multiplication by $h_\phi^2/(2v_\phi^2)$ gives
the corresponding quartic interactions. Their coefficients are
fixed by the covariant kinetic term and the phase rotation, rather
than additional free couplings. In particular, the coefficient of
$h_\phi B_\mu\partial^\mu\chi$ has one power of the portal
coupling because $\mu_B=e_Bv_\phi$.

The potential and derivative terms in Eq.~\eqref{eq:hphichichi}
both contribute to $h_\phi\to\chi\chi$. Denote their coefficients by
\begin{equation}
 g=\frac{m_\chi^2}{v_\phi},\qquad
 a=\frac{M_1^2}{v_\phi M_B^2},\qquad
 \cL_{h_\phi\chi\chi}=-\frac g2h_\phi\chi^2
                         +a h_\phi\partial_\mu\chi\partial^\mu\chi.
 \label{eq:vertex-coefficients}
\end{equation}
The factor $1/2$ in the potential term compensates the two
identical $\chi$ fields when the vertex is taken. The derivative
term has two assignments of the outgoing momenta to the two fields.
With all momenta taken into the vertex, each derivative supplies
its momentum factor, and the vertex is
\begin{equation}
 -\ii\,[g+2a\,k_1\cdot k_2].
 \label{eq:vertex-momenta}
\end{equation}
For the decay, the two outgoing particles satisfy
$k_1^2=k_2^2=m_\chi^2$ and $(k_1+k_2)^2=m_{h_\phi}^2$.
Consequently
\begin{equation}
 2k_1\cdot k_2=m_{h_\phi}^2-2m_\chi^2,
 \qquad
 \mathcal G_{h_\phi\chi\chi}
      =g+a(m_{h_\phi}^2-2m_\chi^2),
 \label{eq:vertex-onshell-step}
\end{equation}
which gives Eq.~\eqref{eq:hphi-amplitude}. The relative sign
of the two contributions follows from the original Lagrangian.
Adding their separate decay probabilities would omit their
interference.

In the rest frame of $h_\phi$, each daughter has momentum magnitude
\begin{equation}
 |\mathbf k|=\frac{m_{h_\phi}}2
            \sqrt{1-\frac{4m_\chi^2}{m_{h_\phi}^2}}.
 \label{eq:two-body-momentum}
\end{equation}
The integrated two-body phase space is
$\int\dd\Phi_2=|\mathbf k|/(4\pi m_{h_\phi})$.
The initial-state normalization contributes $1/(2m_{h_\phi})$,
and the identical final particles contribute $1/2!$. Hence
\begin{equation}
 \Gamma=\frac{1}{2m_{h_\phi}}\frac{1}{2!}
              |\mathcal G_{h_\phi\chi\chi}|^2\int\dd\Phi_2
 =\frac{|\mathcal G_{h_\phi\chi\chi}|^2}
              {32\pi m_{h_\phi}}
                 \sqrt{1-\frac{4m_\chi^2}{m_{h_\phi}^2}},
 \label{eq:width-normalization}
\end{equation}
as used above. The factor of two from identical fields in the
vertex and the factor $1/2!$ in final-state phase space have
different roles; retaining both produces the stated normalization.

The derivative term remains present when the instanton coefficient
is small. Suppressing the mass of $\chi$ therefore does not
automatically suppress this hidden radial decay.

The fermion interactions follow from the Yukawa term in
Eq.~\eqref{eq:yukawa}. For $j=F,\Psi$, unitary gauge gives
\begin{equation}
 \cL_{Y,j}=-m_j\left(1+\frac{h_\phi}{\vp}\right)
 \bar j\exp\!\left[-\ii\gamma_5\frac{M_1\chi}{\vp M_B}\right]j.
 \label{eq:fermion-mass-phase}
\end{equation}
Expanding in the physical fields gives
\begin{align}
 \cL_{Y,j}={}&-m_j\bar j j-\frac{m_j}{\vp}h_\phi\bar j j
       +\ii y_{\chi j}\chi\bar j\gamma_5j
       +\ii\frac{y_{\chi j}}{\vp}h_\phi\chi\bar j\gamma_5j
       +\frac{y_{\chi j}^2}{2m_j}\chi^2\bar j j+\cdots,
       \nonumber\\
 y_{\chi j}&=\frac{m_jM_1}{\vp M_B}
             =\frac{m_j\sigma_\chi}{\vp^2}.
 \label{eq:fermion-couplings}
\end{align}
The appearance of $\gamma_5$ can be derived directly from the
left- and right-handed Yukawa interaction. For one hidden species,
the term and its Hermitian conjugate are
\begin{equation}
 -y_j\phi\bar j_Lj_R-y_j\phi^*\bar j_Rj_L
 =-m_j\left(1+\frac{h_\phi}{v_\phi}\right)
 \bar j\left(e^{\ii\eta/v_\phi}P_R
             +e^{-\ii\eta/v_\phi}P_L\right)j.
 \label{eq:yukawa-projectors}
\end{equation}
Using $P_R+P_L=1$ and $P_R-P_L=\gamma_5$, the matrix
in parentheses is
\begin{equation}
 e^{\ii\eta/v_\phi}P_R+e^{-\ii\eta/v_\phi}P_L
 =\cos(\eta/v_\phi)+\ii\gamma_5\sin(\eta/v_\phi)
 =e^{\ii\gamma_5\eta/v_\phi}.
 \label{eq:yukawa-gammafive}
\end{equation}
In unitary gauge, $\eta=-M_1\chi/M_B$, which yields
Eq.~\eqref{eq:fermion-mass-phase}. Expanding the exponential
to second order uses $\gamma_5^2=1$ and gives
\begin{equation}
 e^{-\ii\gamma_5M_1\chi/(v_\phi M_B)}
 =1-\ii\gamma_5\frac{M_1\chi}{v_\phi M_B}
    -\frac12\left(\frac{M_1\chi}{v_\phi M_B}\right)^2
    +\cdots.
 \label{eq:yukawa-series}
\end{equation}
Multiplication by the overall negative mass term produces the
positive $\ii y_{\chi j}\chi\bar j\gamma_5j$ coupling and
the positive quadratic contact term in
Eq.~\eqref{eq:fermion-couplings}. Multiplying the linear term
by $h_\phi/v_\phi$ gives the mixed hidden contact vertex.
Thus the coupling of $\chi$ to the fermions is determined by
their mass and the projection of the hidden Higgs phase onto
$\chi$. It is not obtained by replacing the Standard Model
Higgs with a new field, and it does not require mixing of the
two radial Higgs excitations.

The scalar, pseudoscalar and contact vertices are thus related by
the same Yukawa interaction. The coupling of $h_\phi$ to a
hidden fermion is $m_j/\vp$; the coupling of the Standard Model
Higgs $h_H$ to an ordinary fermion is $m_f/v_H$. They belong
to separate radial sectors.

If a hidden fermion-pair channel is kinematically open, the
pseudoscalar coupling gives
\begin{equation}
 \Gamma(\chi\to j\bar j)=
 \frac{N_j y_{\chi j}^2m_\chi}{8\pi}
       \sqrt{1-\frac{4m_j^2}{m_\chi^2}},\qquad
 N_\Psi=1,\quad N_F=2.
 \label{eq:fermion-width}
\end{equation}
For $j=F$ this is only a leading partonic expression. The unbroken
hidden group confines, so observable final states are hidden hadrons.
An inclusive partonic approximation requires energies well above
the strong scale and away from hadronic thresholds; it is not a
formula for free asymptotic $F$ particles. The light illustration
below has both fermion-pair channels closed. Its lifetime through
off-shell gauge fields and possible hidden bound states is not
computed here. The local gauge vertices in
Eq.~\eqref{eq:physical-wz} must be combined with the corresponding
fermion-loop contributions in physical amplitudes.

\section{Potential-coefficient dependence and an illustrative spectrum}
\label{sec:mass-plots}

\begin{figure}[tbp]
 \centering
 \includegraphics[width=0.88\textwidth]{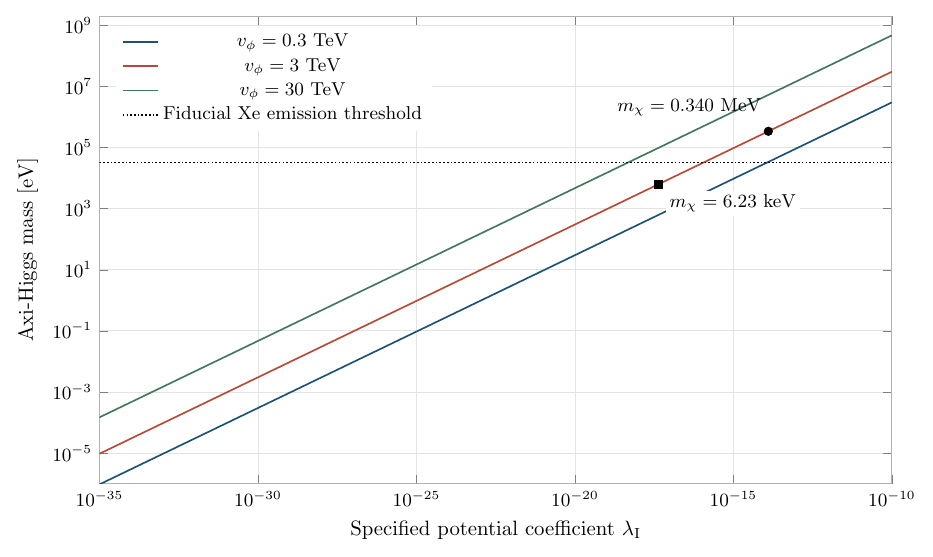}
 \caption{Mass as a function of the specified potential coefficient,
 for $v_\phi=0.3$, $3$ and $30$ TeV, $M_1=10$ TeV and $e_B=0.40$.
 The points mark the $0.340$ MeV spectrum illustration and the
 $6.23$ keV emission example. The horizontal line is the maximum
 axi-Higgs mass, $32.6$ keV, that can be emitted for
 $m_\Psi=1.10$ TeV, a representative xenon nuclear mass and the
 incident speed $v=7.7\times10^{-4}c$: emission requires a mass
 below this line for that chosen collision. No halo abundance
 or microscopic matching is assumed.}
 \label{fig:mass-action}
\end{figure}

\begin{figure}[tbp]
 \centering
 \includegraphics[width=0.88\textwidth]{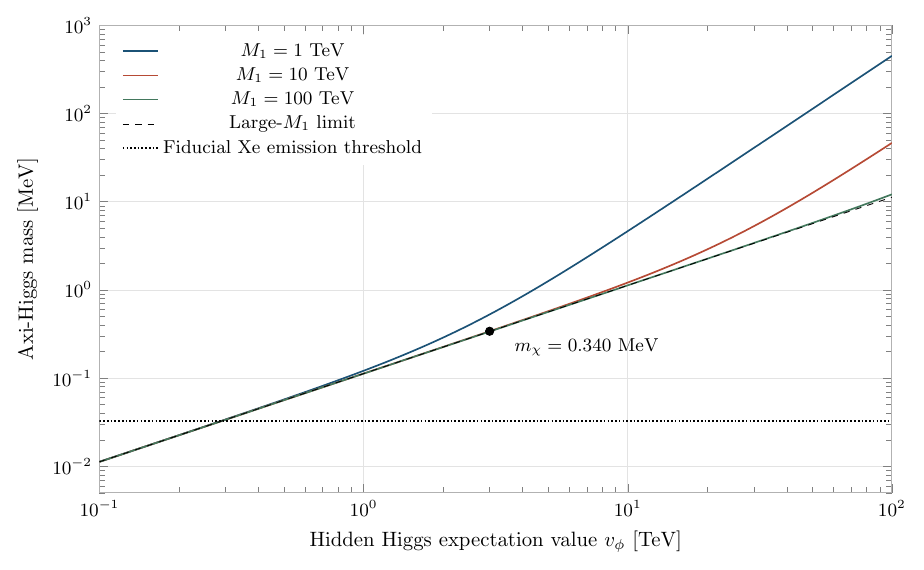}
 \caption{Mass versus the hidden Higgs expectation value for
 $M_1=1$, $10$ and $100$ TeV, $e_B=0.40$ and the specified
 coefficient $\lambda_{\mathrm I}=1.2664\times10^{-14}$.
 The dashed curve is $m_\chi=\sqrt{\lambda_{\mathrm I}}v_\phi$.
 The horizontal dotted line gives the same $32.6$ keV maximum emitted
 mass for the incident speed used in Fig.~\ref{fig:mass-action}.
 The marked point is above this kinematic production boundary.}
 \label{fig:mass-vev}
\end{figure}

The mass relation in Eq.~\eqref{eq:instanton-mass} displays the
consequences of the specified potential independently of a microscopic
calculation of its coefficient. Figure~\ref{fig:mass-action} scans
$\lambda_{\mathrm I}$, and Fig.~\ref{fig:mass-vev} varies the hidden
Higgs scale at fixed coefficient. These are conditional effective-model
curves. They are not predictions obtained by scanning $g_X$ in the
confining theory of Section~\ref{sec:hidden-dynamics}.

At fixed $v_\phi,M_1,e_B$, reducing $\lambda_{\mathrm I}$ by
a factor of one hundred reduces $m_\chi$ by ten. For
$v_\phi=3$ TeV, $M_1=10$ TeV and $e_B=0.40$, the coefficients
$3.78\times10^{-11}$, $1.27\times10^{-14}$ and
$4.25\times10^{-18}$ give masses of approximately
$18.6$ MeV, $0.340$ MeV and $6.23$ keV.

For $M_1\gg e_Bv_\phi$, the mass approaches
$\sqrt{\lambda_{\mathrm I}}v_\phi$. When $M_1\ll e_Bv_\phi$,
the normalization supplies an additional factor proportional to
$v_\phi/M_1$. The second plot holds the same coefficient fixed
while comparing these limits. For a hidden radial mass of $1.50$
TeV the largest mass in these two displayed scans is below $0.5$ GeV;
the mass hierarchy supporting slow fixed-radial-field configurations
is therefore satisfied throughout. It does not justify arbitrary
high-frequency configurations or the assumed nonperturbative matching.

For an illustrative effective-potential spectrum we choose
\begin{gather}
 v_H=246.22~\mathrm{GeV},\qquad m_{h_H}=125.25~\mathrm{GeV},
 \qquad \vp=3~\mathrm{TeV},\qquad m_{h_\phi}=1.50~\mathrm{TeV},\nonumber\\
 M_1=10~\mathrm{TeV},\qquad g_B=0.20,\qquad
 \lambda_{\mathrm I}=1.2664\times10^{-14},\nonumber\\
 m_\Psi=1.10~\mathrm{TeV},\qquad m_F=0.600~\mathrm{TeV}.
 \label{eq:inputs}
\end{gather}
The resulting vector mass, phase scale and conditional pseudoscalar
mass are
\begin{equation}
 M_B\simeq10.072~\mathrm{TeV},\qquad
 \sigma_\chi\simeq2.979~\mathrm{TeV},\qquad
 m_\chi\simeq0.340~\mathrm{MeV}.
 \label{eq:benchmark-mass}
\end{equation}
To relate this example to a possible confining scale, parameterize
the normalization of the angular curvature at its minimum as
\begin{equation}
 \lambda_{\mathrm I}v_\phi^4=C_{\rm np}\Lambda_X^4,
 \qquad
 \Lambda_X=v_\phi\left(\frac{\lambda_{\mathrm I}}{C_{\rm np}}\right)^{1/4},
 \qquad
 m_\chi=\frac{\sqrt{C_{\rm np}}\Lambda_X^2}{\sigma_\chi}.
 \label{eq:curvature-scale}
\end{equation}
Here $C_{\rm np}>0$ is an undetermined dimensionless nonperturbative
normalization, defined using the one-loop scale in
Eq.~\eqref{eq:hidden-scale}. This relation fixes the quadratic term;
it does not establish a cosine over the entire period or the assumed
linear radial dependence. Inverting the same running formula gives
\begin{equation}
 \frac{8\pi^2}{g_X^2(v_\phi)}
 =\frac{22}{3}\ln\frac{m_F}{\Lambda_X}
       -\frac{20}{3}\ln\frac{m_F}{v_\phi},
 \qquad \alpha_X=\frac{g_X^2}{4\pi}.
 \label{eq:hidden-scale-inverse}
\end{equation}
For the illustrative choice $C_{\rm np}=1$, the $0.340$ MeV
example corresponds to $\Lambda_X\simeq1.006$ GeV,
$8\pi^2/g_X^2(v_\phi)\simeq57.59$ and
$\alpha_X(v_\phi)\simeq0.1091$. Likewise,
$\lambda_{\mathrm I}=4.25\times10^{-18}$ gives $m_\chi\simeq6.23$
keV, $\Lambda_X\simeq0.1362$ GeV and
$\alpha_X(v_\phi)\simeq0.08695$. Thus confinement can accommodate
the scale of either mass with a different coupling from the
$\alpha_X=0.196$ example. These are conditional one-loop estimates,
not four-digit predictions of nonperturbative dynamics. An unknown
$C_{\rm np}$ changes the inferred scale and coupling, but their
sensitivities differ. At fixed $\lambda_{\mathrm I}$,
$\Lambda_X\propto C_{\rm np}^{-1/4}$ and
\begin{equation}
 \frac{8\pi^2}{g_X^2(v_\phi;C_{\rm np})}
 =\left.\frac{8\pi^2}{g_X^2(v_\phi)}\right|_{C_{\rm np}=1}
       +\frac{11}{6}\ln C_{\rm np}.
 \label{eq:normalization-sensitivity}
\end{equation}
For the $0.340$ MeV example, varying $C_{\rm np}$ from $10^{-3}$
to $10^3$ moves $\Lambda_X$ from $5.66$ to $0.179$ GeV, while
$\alpha_X(3~\mathrm{TeV})$ changes only from $0.140$ to $0.0894$.
Thus the inferred coupling remains of order $0.1$ over this wide
normalization range, within the same one-loop and heavy-fermion
approximation. The coefficient
is the assumed total angular curvature, not a small term to be added
while ignoring an existing confining contribution.

The radial quartics are
\begin{equation}
 \lambda_H=\frac{m_{h_H}^2}{2v_H^2}\simeq0.129,\qquad
 \lambda_\phi=\frac12\left(\frac{m_{h_\phi}^2}{\vp^2}
                              -\lambda_{\mathrm I}\right)\simeq0.125.
 \label{eq:benchmark-quartics}
\end{equation}
The parent-scalar values are $w=M_1/e_B=25$ TeV and, for
$\lambda_S=1$, $m_\rho\simeq35.4$ TeV. The Yukawa couplings
are $y_\Psi\simeq0.519$, $y_F\simeq0.283$,
$y_{\chi\Psi}\simeq0.364$ and $y_{\chi F}\simeq0.199$.
For the chosen linear radial operator,
Eq.~\eqref{eq:hphi-width} gives
$\Gamma(h_\phi\to\chi\chi)\simeq3.63$ GeV. The derivative
vertex follows from the kinetic action, whereas the potential
part of this width depends on the radial profile. The comparison
in Eq.~\eqref{eq:radial-power-vertex} shows why changing its power
from $1$ to $4/11$ leaves the quoted width unchanged at the displayed
precision when the physical masses are fixed.

Stability of $\Psi$ does not fix its abundance, and
the closed two-fermion decay channels do not determine the lifetime
of $\chi$. Cancellation of visible gauge anomalies is not a proof
of its absolute stability: the hidden gauge vertices and their
couplings through virtual portal fields still require a decay analysis.
In a confining completion, hidden glueballs and hadrons
must also be included; dark-glueball cosmology illustrates the
importance of their production and decay history~\cite{Forestell2018}.
Their abundances, lifetimes and decay products would be needed before
testing light-element constraints~\cite{Pitrou2018,Kawasaki2018}
or radiation-density constraints~\cite{Planck2018}. None is inferred from the particle masses alone.

Suppose $\chi$ attained chemical equilibrium with the visible plasma,
decoupled while relativistic and survived to the present without
further depletion or additional entropy dilution. For a real scalar,
$n_\chi/s=45\zeta(3)/(2\pi^4g_{*s,d})$, where $n_\chi$ is its
number density, $s$ the entropy density and $g_{*s,d}$ the effective
entropy degrees of freedom at decoupling~\cite{PDGDM2025}.
Taking $g_{*s,d}=106.75$ illustratively gives
$\Omega_\chi h_0^2\simeq243$ for $m_\chi=0.340$ MeV and
$4.4$ for $m_\chi=6.23$ keV, with
$h_0=H_0/(100~\mathrm{km\,s^{-1}\,Mpc^{-1}})$.
Both exceed the Planck base-$\Lambda$CDM determination
$\Omega_c h_0^2=0.120\pm0.001$~\cite{Planck2018}.
A history preventing equilibration, sufficient late entropy dilution,
or suitable depletion or decay would therefore have to be established
to make these examples viable. The dependence of relic production on
reheating and entropy release is discussed in Ref.~\cite{Giudice2001}.
This estimate assumes an equilibrium
population; the portal and Yukawa couplings alone do not determine
its decoupling history, and a relativistic scattering-rate estimate
cannot be extrapolated below the heavy-particle thresholds without
including their suppressed populations and the appropriate thermal
averages~\cite{GondoloGelmini1991}. No such history or lifetime
is calculated here, and no temperature dependence is introduced into
the potential. Whether a reheating history can suppress the $\chi$
population while supplying the incident $\Psi$ population used below
remains open and requires their coupled production and evolution;
no allowed reheating window is established by the rate estimates alone.

\section{Elastic nuclear scattering and the axi-Higgs emission threshold}
\label{sec:scattering}

The stable singlet $\Psi$ permits a coherent elastic-scattering
calculation and a specification of the threshold for radiative
scattering through the same vector portal. We give the elastic
differential cross section; for emission we identify the tree diagrams
and kinematic requirements without calculating the three-body rate. In the process considered here, $\Psi$ is the
incident particle, the nucleus recoils, and $\chi$ may be emitted
during the collision. Neither Higgs mediates a tree-level interaction
between $\Psi$ and ordinary matter, since $h_H$ and $h_\phi$
do not mix. The exchange between the two sectors is the vector $B_\mu$.
Appendix~\ref{app:dark-matter-kinematics} derives the exact relativistic
production threshold, its reduced-mass limit, the relation to Galactic
speed distributions, and the interpretation of the horizontal lines
in Figs.~\ref{fig:mass-action} and~\ref{fig:mass-vev}.
For the emission example we use $\lambda_{\mathrm I}=4.25\times10^{-18}$,
so that $m_\chi=6.23$ keV, retaining the other inputs in
Eq.~\eqref{eq:inputs}. This opens the channel at the reference speed;
the heavier spectrum example provides a useful threshold comparison.

The portal charges of $\Psi$ imply
\begin{equation}
 \cL_{\Psi B}=-g_BB_\mu\bar\Psi\gamma^\mu\gamma_5\Psi,
 \qquad g_V=\frac{g_B}{2}(z_L+z_R)=0,
 \qquad g_A=\frac{g_B}{2}(z_R-z_L)=-g_B.
 \label{eq:VA}
\end{equation}
Ordinary matter couples through the vector $B-L$ current of
Eq.~\eqref{eq:visible-current}. Thus the dark fermion current is
axial while the nuclear current is vectorial.

Let the incident and outgoing singlet momenta be $p,p'$ and the
nuclear momenta be $P,P'$. For a nucleus of mass $m_A$ and baryon
number $A$, define $q=P'-P$, $t=q^2$ and $J=P+P'$.
In the leading coherent, spin-zero approximation its vector current
is $AF_V(t)J^\mu$, with $F_V(0)=1$. The factor $A$ accounts
for the common nucleon charge; the form factor describes the loss
of coherence at finite momentum transfer. A Helm form factor is
a standard approximation for this purpose~\cite{Helm:1956,LewinSmith1996}.

The elastic amplitude is
\begin{equation}
 \mathcal M_B^{\rm el}=
 \frac{g_BAF_V(t)g_A}{t-M_B^2}
           \bar u(p')\slashed J\gamma_5u(p).
 \label{eq:elastic-amplitude}
\end{equation}
The longitudinal vector numerator drops out because
$q\cdot J=P'^2-P^2=0$. In the nuclear rest frame, the recoil
energy $E_R$ and incident energy $E$ give
\begin{align}
 t&=-2m_AE_R,\qquad J^2=4m_A^2-t,\qquad
 p\cdot J=2m_AE+t/2,\nonumber\\
 \overline{|\mathcal M_B^{\rm el}|^2}
 &=\left|\frac{g_BAF_V(t)g_A}{t-M_B^2}\right|^2
   \left[4(p\cdot J)^2+(t-4m_\Psi^2)J^2\right],\nonumber\\
 \frac{\dd\sigma_{\rm el}}{\dd E_R}
 &=\frac{\overline{|\mathcal M_B^{\rm el}|^2}}
              {32\pi m_Ap_{\rm lab}^2},
 \qquad p_{\rm lab}^2=E^2-m_\Psi^2.
 \label{eq:elastic-cross}
\end{align}
The overline denotes a sum over final $\Psi$ spins and an average
over its two incident spin states; the spin-zero nuclear approximation
requires no further spin average. The axial--vector amplitude is
suppressed at low speed. There is
no scalar-exchange term to add in the present model.

\begin{figure}[!t]
\centering
\begin{tikzpicture}[
 fermion/.style={thick,-{Latex[length=1.8mm]}},
 boson/.style={thick,decorate,decoration={snake,amplitude=1pt,segment length=4.5pt}},
 scalar/.style={thick,dashed},every node/.style={font=\small}]
 \begin{scope}
  \draw[fermion] (-1.8,1.5)--(-.65,1.5);
  \draw[fermion] (-.65,1.5)--(0,1.5);
  \draw[fermion] (0,1.5)--(1.8,1.5);
  \draw[fermion] (-1.8,0)--(0,0);
  \draw[fermion] (0,0)--(1.8,0);
  \draw[boson] (0,0)--(0,1.5) node[midway,right] {$B$};
  \draw[scalar] (-.65,1.5)--(-1.1,2.35) node[above] {$\chi$};
  \node[left] at (-1.8,1.5) {$\Psi$};\node[left] at (-1.8,0) {$N$};
  \fill (0,0) circle (1.5pt);\fill (0,1.5) circle (1.5pt);\fill (-.65,1.5) circle (1.5pt);
  \node at (0,-.5) {Initial-line emission};
 \end{scope}
 \begin{scope}[xshift=7cm]
  \draw[fermion] (-1.8,1.5)--(0,1.5);
  \draw[fermion] (0,1.5)--(.65,1.5);
  \draw[fermion] (.65,1.5)--(1.8,1.5);
  \draw[fermion] (-1.8,0)--(0,0);
  \draw[fermion] (0,0)--(1.8,0);
  \draw[boson] (0,0)--(0,1.5) node[midway,right] {$B$};
  \draw[scalar] (.65,1.5)--(1.1,2.35) node[above] {$\chi$};
  \node[left] at (-1.8,1.5) {$\Psi$};\node[left] at (-1.8,0) {$N$};
  \fill (0,0) circle (1.5pt);\fill (0,1.5) circle (1.5pt);\fill (.65,1.5) circle (1.5pt);
  \node at (0,-.5) {Final-line emission};
 \end{scope}
\end{tikzpicture}
\caption{The two fermion-leg contributions to axi-Higgs radiation
with vector exchange. The incident state is the stable singlet $\Psi$;
the nucleus $N$ receives momentum through the virtual portal vector,
and $\chi$ is emitted. These two diagrams interfere. The hidden $h_\phi B\chi$
vertex supplies an additional internal-emission contribution,
described in the text.}
\label{fig:nuclear-radiation}
\end{figure}
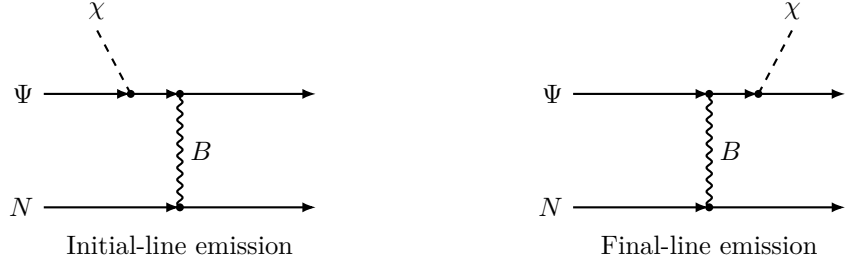

The same vector exchange allows the radiative process
\begin{equation}
 \Psi(p)+N(P)\longrightarrow\Psi(p')+N(P')+\chi(k).
 \label{eq:radiation}
\end{equation}
Figure~\ref{fig:nuclear-radiation} shows emission from the incident
and outgoing singlet lines, using $y_{\chi\Psi}$ in
Eq.~\eqref{eq:fermion-couplings}. The action also gives an
internal-emission diagram: the singlet emits a virtual $h_\phi$,
which meets the exchanged $B$ and produces $\chi$ through
Eq.~\eqref{eq:hphiBchi}. This is a hidden radial interaction, not
a coupling to the Standard Model Higgs. All three contributions
must be added before squaring the tree amplitude.

The exact threshold condition for axi-Higgs emission is
\begin{equation}
 s=m_\Psi^2+m_A^2+2m_AE
          \geq(m_\Psi+m_A+m_\chi)^2.
 \label{eq:threshold}
\end{equation}
For nonrelativistic incident particles and $m_\chi\ll m_\Psi,m_A$,
this becomes
\begin{equation}
 \frac12\mu_{\Psi A}v_{\rm rel}^2\geq m_\chi,
 \qquad \mu_{\Psi A}=\frac{m_\Psi m_A}{m_\Psi+m_A}.
 \label{eq:nonrel-threshold}
\end{equation}
For a xenon nucleus approximated by $m_A=122$ GeV and
$m_\Psi=1.10$ TeV, the reduced mass is $109.8$ GeV. At
$v_{\rm rel}=7.7\times10^{-4}$ the available energy is only
$32.6$ keV, so emission of the $6.23$ keV axi-Higgs is kinematically
allowed. Its minimum incident speed is approximately
$101~\mathrm{km\,s^{-1}}$. By contrast, the $0.340$ MeV spectrum
illustration cannot radiate at the reference speed. It requires
\begin{equation}
 v_{\rm rel}\geq\sqrt{\frac{2m_\chi}{\mu_{\Psi A}}}
       \simeq2.49\times10^{-3}\simeq746~\mathrm{km\,s^{-1}}.
 \label{eq:emission-speed}
\end{equation}
Such speeds can occur only in the extreme laboratory-frame tail of
some halo distributions; accessibility and suppression depend on
the distribution and the Earth's motion~\cite{Baxter2021,LavalleMagni2015}.
The lighter example avoids this tail requirement, but kinematic
accessibility alone does not determine an observable rate.

A radiative cross section would require the sum of all three
amplitudes described above and the three-body phase-space integral.
An event rate would additionally require the incident distribution
and detector response~\cite{LewinSmith1996,Baxter2021}.
Neither is calculated here, and the
threshold calculation is not a sensitivity forecast.

\section{Conclusions}

We have developed a one-hidden-Higgs effective realization of the
axi-Higgs mechanism, with one Standard Model doublet $H$ and one
charged hidden scalar $\phi$. Their scalar potentials remain separate.
The portal vector couples to the anomaly-free visible $B-L$ current
and to chiral hidden matter, whose anomalous variation is canceled
by Wess--Zumino terms. The Abelian A--B example makes the vector
Ward identities and their local completion explicit before the
non-Abelian hidden representation is introduced.

The scalar origin of the Stueckelberg kinetic term and the physical
phase can be derived without specifying the periodic energy. Removing
the heavy parent radial mode leaves $b$ with $M_1=e_Bw$.
The retained hidden Higgs phase and $b$ then form the eaten
Goldstone and a gauge-invariant field $\chi$, whose normalization
is $\sigma_\chi=v_\phi M_1/M_B$. These results follow from the
kinetic action and do not depend on an instanton estimate.

For the specified leading periodic operator, the mass is
$m_\chi^2=\lambda_{\mathrm I}v_\phi^4/\sigma_\chi^2$.
We have derived its radial vertices, their interference with the
derivative interaction in $h_\phi\to\chi\chi$, and the associated
Yukawa vertices. Comparing radial powers $p=1$ and $p=4/11$ at
fixed physical masses leaves the illustrative radial width unchanged
at the quoted precision because its derivative contribution dominates.
The fixed-radial-field limit gives the sine-Gordon
equation and its planar kink. In the compact parent realization the
first harmonic has one inequivalent minimum per physical period;
the kink calculation alone is not a prediction of a stable
cosmological wall network.

The unbroken $SU(2)_X$
theory is not cut off in the infrared by the singlet Higgs expectation
value. One-loop threshold matching shows that the illustrative
coupling with $8\pi^2/g_X^2(v_\phi)=32$ runs to a strong scale of
about $33$ GeV. A tiny potential coefficient cannot be inferred by
retaining only its instanton exponential at $v_\phi$. Heavy-fermion
matching also shows that the strong contribution need not have the
linear radial dependence assumed in the retained operator. Conversely,
with unit nonperturbative normalization, a strong scale near $1$ GeV
and $\alpha_X(3~\mathrm{TeV})\simeq0.109$ accommodate the scale of
the $0.340$ MeV example; the $6.23$ keV example corresponds to about
$0.136$ GeV and $\alpha_X\simeq0.087$. These estimates connect the
light masses to possible gauge-sector parameters without claiming a
matched angular shape or radial dependence. The imposed relative-phase
symmetry has the hidden Peccei--Quinn interpretation, with
$f_\chi=\sigma_\chi$. It protects against an unrestricted perturbative
cosine but leaves the nonperturbative matching requirement.

The theta-vacuum appendix identifies the invariant angular variable
by removing the Yukawa phase and including the anomalous Jacobian.
An instanton--anti-instanton sum explains a cosine in the dilute
regime and relates it to the potential used in the scalar calculation.
That argument fixes the phase convention, but does not calculate
the amplitude or establish the dilute regime for the stated matter
content. A quantitative determination of its full vacuum energy
and radial dependence remains necessary for a microscopic prediction.

Vector exchange gives an explicit coherent elastic nuclear
cross section. We have identified the three contributions required
for additional axi-Higgs emission and evaluated its threshold.
The $6.23$ keV emission example is open at the reference speed on
xenon. The heavier $0.340$ MeV example instead requires a speed near
$746~\mathrm{km\,s^{-1}}$ and can be accessible only in an extreme
halo tail. For chemical equilibrium with the visible plasma followed
by relativistic decoupling with $g_{*s,d}=106.75$, survival to the
present and no later depletion or additional entropy dilution, the
conditional estimate gives $\Omega_\chi h_0^2\simeq243$ and $4.4$
for the $0.340$ MeV and $6.23$ keV examples, respectively, both
above the Planck base-$\Lambda$CDM value near
$0.12$~\cite{Planck2018}. These examples therefore
require a history that avoids the assumed equilibrium population or
provides sufficient dilution, depletion or decay; a compatible
production history for the incident $\Psi$ population remains to be
established. The emission rate and the actual relic populations have
not been calculated. The established results are the gauge completion,
the physical phase, and the particle dynamics conditional on the
specified periodic interaction; a confining completion must supply
the additional nonperturbative information.

\section*{Acknowledgments}

This work is partially supported by INFN,
Iniziativa Specifica \emph{QG-SKY}.

\appendix
\section{Gauge identities and the visible current}
\label{app:gauge-identities}

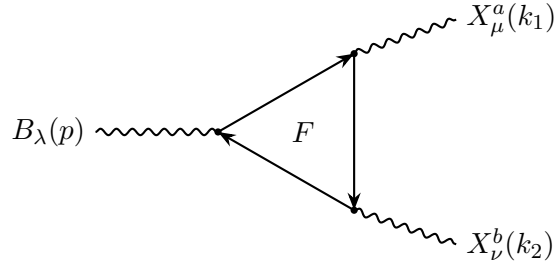
\begin{figure}[htbp]
 \centering
 \begin{tikzpicture}[>=Stealth,scale=0.9]
  \coordinate (A) at (0,0);
  \coordinate (C) at (2,1.15);
  \coordinate (D) at (2,-1.15);
  \draw[thick,->] (A)--(C);
  \draw[thick,->] (C)--(D);
  \draw[thick,->] (D)--(A);
  \draw[thick,decorate,decoration={snake,amplitude=1.2pt,segment length=6pt}]
       (-1.8,0)--(A);
  \draw[thick,decorate,decoration={snake,amplitude=1.2pt,segment length=6pt}]
       (C)--(3.5,1.65);
  \draw[thick,decorate,decoration={snake,amplitude=1.2pt,segment length=6pt}]
       (D)--(3.5,-1.65);
  \fill (A) circle (1.5pt);
  \fill (C) circle (1.5pt);
  \fill (D) circle (1.5pt);
  \node[left] at (-1.8,0) {$B_\lambda(p)$};
  \node[right] at (3.5,1.65) {$X_\mu^a(k_1)$};
  \node[right] at (3.5,-1.65) {$X_\nu^b(k_2)$};
  \node at (1.25,0) {$F$};
 \end{tikzpicture}
 \caption{One orientation of the hidden-fermion $BXX$ triangle.
 The loop contains the Dirac doublet $F$. Its portal current is
 axial, $J_{B,F}^\lambda=-\bar F\gamma^\lambda\gamma_5F$,
 while its two hidden-color currents are vectorial. The second
 fermion-loop orientation is also included in the amplitude.
 This triangle supplies the mixed anomaly canceled in
 Fig.~\ref{fig:BXX-WZ-cancellation}; the stable color singlet
 $\Psi$ does not contribute to it.}
 \label{fig:lh-BXX-triangle}
\end{figure}

The hidden Dirac doublet is vectorlike under $SU(2)_X$ and chiral
under the portal. The two vertices, with their gauge couplings
factored out, are
\begin{equation}
 \gamma^\lambda(z_LP_L+z_RP_R)=-\gamma^\lambda\gamma_5,
 \qquad \gamma^\mu T^a,\qquad (z_L,z_R)=(+1,-1).
 \label{eq:triangle-currents}
\end{equation}
Thus Fig.~\ref{fig:lh-BXX-triangle} has one axial and two vector
insertions. The corresponding chiral trace is
$(z_L-z_R)\operatorname{tr}(T^aT^b)=\delta^{ab}$,
which is the coefficient $C_{BXX}=1$ in
Eq.~\eqref{eq:anomaly-traces}. Keeping the hidden vector identities
fixes the placement of the mixed anomaly on the portal leg.

In the baseline spectrum the summed visible
fermions are anomaly free and their $B-L$ vector current is conserved.
At the hidden end, gauge invariance involves the fermion triangle,
the shifting scalar phases and the local Wess--Zumino term together.
The completed vertices consequently satisfy
\begin{equation}
 p_\lambda\Gamma_{BXX}^{\lambda\alpha\beta}
 -M_1\Gamma_{bXX}^{\alpha\beta}
 -\mu_B\Gamma_{\eta XX}^{\alpha\beta}=0,
 \qquad
 p_\lambda\Gamma_{BXX}^{\lambda\alpha\beta}
 -M_B\Gamma_{GXX}^{\alpha\beta}=0.
 \label{eq:completed-identity}
\end{equation}
Common Fourier vertex phases are understood in this identity.
The $bXX$ term cancels the anomalous part of the fermion divergence;
the $\eta XX$ term accounts for its mass insertion. The rotation
combines the two shifting phases into $G$. Since $\chi$ is invariant,
it does not provide a gauge variation. An isolated massive-fermion
triangle consequently need not be transverse in the portal index
for the completed theory to be gauge consistent \cite{CIM1,CIM2}.
The ordinary nuclear vector-current matrix element is conserved
at the order considered here.

In an $R_\xi$ gauge the vector propagator is
\begin{equation}
 D_{\mu\nu}^{(\xi)}(q)=
 \frac{-\ii P^T_{\mu\nu}}{q^2-M_B^2+\ii0}
 +\frac{-\ii\xi P^L_{\mu\nu}}{q^2-\xi M_B^2+\ii0},
 \quad P^L_{\mu\nu}=\frac{q_\mu q_\nu}{q^2},\quad
 P^T_{\mu\nu}=g_{\mu\nu}-P^L_{\mu\nu}.
 \label{eq:propagator}
\end{equation}
For a conserved visible matrix element $J_{\rm vis}$,
$q\cdot J_{\rm vis}=0$ removes the second term. If
$\mathcal H^\nu$ denotes the completed hidden current matrix
element, the exchange is proportional to
\begin{equation}
 \mathcal M_{\rm vis\to hid}=
 \frac{-\ii g_B}{q^2-M_B^2+\ii0}
 J_{{\rm vis},\nu}P_T^{\nu\rho}\mathcal H_\rho.
 \label{eq:visible-projection}
\end{equation}
The longitudinal anomaly-pole structure is discussed in Ref.~\cite{CLM}.
A purely longitudinal contribution
does not survive this contraction. The physical $\chi$ interactions
derived from the kinetic and potential terms remain present.
Current conservation removes a component of vector exchange;
it does not remove the physical pseudoscalar from the spectrum.

\section{The Yukawa phase and the theta-vacuum energy}
\label{app:theta-vacuum}

A chiral change of fermion variables removes the phase from the
Yukawa mass but transfers it to the coefficient of the topological
density. The Wess--Zumino term then makes that coefficient gauge
invariant. We give this connection for the hidden $SU(2)_X$ theory,
keeping the radial field constant when discussing its angular
energy. The relation between the axial anomaly and pseudoscalar
vacuum dynamics has its classic formulation in
Refs.~\cite{Veneziano1979,DiVecchiaVeneziano1980}.

With the density $Q_X$ defined in Eq.~\eqref{eq:hidden-density},
introduce a fixed, dimensionless parameter $\theta_X$ through
\begin{equation}
 \cL_{\theta,X}=\theta_X Q_X,\qquad
 \nu=\int\dd^4x_E\,Q_{X,E}\in\mathbb Z,\qquad
 Z_X(\theta_X)=\sum_{\nu\in\mathbb Z}e^{\ii\nu\theta_X}Z_{X,\nu}.
 \label{eq:alt-theta-sectors}
\end{equation}
Here $x_E$ denotes Euclidean coordinates, $\nu$ is the integer
topological charge for finite-action configurations with the usual
boundary conditions, and $Z_{X,\nu}$ is the functional integral
restricted to that sector. The relative phase $e^{\ii\nu\theta_X}$
defines the theta-vacuum weighting. In particular, $\theta_X$ and
$\theta_X+2\pi$ give the same weights. The angle $\theta_X$ is a
parameter of the action, not another dynamical scalar.

For a fixed magnitude $r=v_\phi+h_\phi>0$, write
$\phi=r e^{\ii\beta}/\sqrt2$, where $\beta=\eta/v_\phi$.
Choosing $y_F$ real and positive, the Yukawa term is
\begin{equation}
 \cL_{m,F}=-m_F(r)\bar F e^{\ii\gamma_5\beta}F,
 \qquad m_F(r)=\frac{y_Fr}{\sqrt2}.
 \label{eq:alt-mass-phase}
\end{equation}
The change of integration variables
\begin{equation}
 F=e^{-\ii\gamma_5\beta/2}F',\qquad
 \bar F=\bar F'e^{-\ii\gamma_5\beta/2}
 \label{eq:alt-chiral-rotation}
\end{equation}
makes this mass real, since the two rotation factors cancel
$e^{\ii\gamma_5\beta}$. The regulated fermion measure also changes.
Its anomalous Jacobian adds the hidden-color term
\begin{equation}
 \Delta\cL_{\rm Jac}=\beta Q_X.
 \label{eq:alt-jacobian}
\end{equation}
This is the convention in which a Dirac mass phase contributes
$+\arg\mathcal M_F$ to the theta angle. Equivalently, the
hidden-color part of the axial identity is
$\partial_\mu j_5^\mu=2\ii m_F\bar F'\gamma_5F'+2Q_X$,
with $j_5^\mu=\bar F'\gamma^\mu\gamma_5F'$~\cite{Fujikawa1979}.
One fundamental Dirac multiplet has $2T(\mathbf2)=1$ in the
mass-phase Jacobian; its two color components are not two flavors.
The singlet $\Psi$ has no hidden-color contribution. Abelian
anomaly terms are part of the same gauge completion, but do not
enter the hidden $SU(2)_X$ theta angle displayed here.

If $\beta$ depends on position, the fermion kinetic term also
acquires $(\partial_\mu\beta)j_5^\mu/2$. This term vanishes for
the constant fields used to evaluate the energy, and is retained
when the rotated variables are used for spacetime-dependent
processes. The rotation is a change of variables in the same
theory; its Jacobian is not a second interaction to add to an
unrotated fermion calculation.

The coefficient $c_X$ in Eq.~\eqref{eq:anomaly} gives
\begin{equation}
 -\frac{b}{4M_1}c_X X^a_{\mu\nu}\widetilde X^{a\mu\nu}
 =-\frac{e_Bb}{M_1}Q_X.
 \label{eq:alt-wz-angle}
\end{equation}
Adding the original angle, the Jacobian and this Wess--Zumino
term therefore yields
\begin{equation}
 \cL_{\theta,X}+\Delta\cL_{\rm Jac}+\cL_{{\rm WZ},X}
 =\vartheta Q_X,\qquad
 \vartheta=\theta_X+\frac{\eta}{v_\phi}-\frac{e_Bb}{M_1}
           =\theta_X-\frac{\chi}{\sigma_\chi}.
 \label{eq:alt-effective-angle}
\end{equation}
The shifts $\delta(\eta/v_\phi)=e_B\omega$ and
$\delta(e_Bb/M_1)=e_B\omega$ cancel. Thus $\vartheta$ is gauge
invariant and the eaten Goldstone does not enter it. The field
$\chi$ here is precisely the canonical coordinate derived from
the kinetic action; no separate background field is required.
For the energy calculation we simply evaluate it at a constant
value.

Let $\mathcal V_4$ be the Euclidean four-volume. The energy density
at fixed $r$ and $\vartheta$ is
\begin{equation}
 E_X(r,\vartheta)=-\lim_{\mathcal V_4\to\infty}
           \frac{1}{\mathcal V_4}\log Z_X(r,\vartheta).
 \label{eq:alt-energy-definition}
\end{equation}
To display the origin of a cosine, retain a dilute gas of
instantons and anti-instantons~\cite{BPST1975,tHooft1976}.
An instanton has charge $+1$ and weight $e^{\ii\vartheta}$;
an anti-instanton has charge $-1$ and the conjugate weight.
Write $\zeta(r)$ for their common one-event weight per unit
four-volume, including the determinant, the size integral and
the factor $e^{-S_{\mathrm I}}$. It has mass dimension four.
Neglecting interactions between these events, their sum gives
\begin{align}
 \frac{Z_X(r,\vartheta)}{Z_{X,\mathrm{pert}}(r)}
 &=\sum_{n_+,n_-=0}^{\infty}
   \frac{[\mathcal V_4\zeta(r)]^{n_++n_-}}{n_+!n_-!}
       e^{\ii(n_+-n_-)\vartheta}\nonumber\\
 &=\exp\!\left[2\mathcal V_4\zeta(r)\cos\vartheta\right],
 \qquad
 E_X(r,\vartheta)=E_{X,0}(r)-2\zeta(r)\cos\vartheta.
 \label{eq:alt-instanton-sum}
\end{align}
Here $Z_{X,\mathrm{pert}}(r)$ denotes the normalization with no instanton
events and $E_{X,0}(r)$ its phase-independent energy. The dilute
gas is an illustration of the leading harmonic; it does not by
itself establish quantitative control of strongly coupled hidden
dynamics.

The operator retained in this paper amounts to the effective
identification
\begin{equation}
 2\zeta(r)=\sqrt2\kappa r=\lambda_{\mathrm I}v_\phi^3r,
 \qquad
 V'_{\theta}=-\left[\kappa e^{\ii\theta_X}
                 \phi e^{-\ii e_Bb/M_1}+\mathrm{h.c.}\right]
             =-\sqrt2\kappa r\cos\vartheta.
 \label{eq:alt-potential-matching}
\end{equation}
For one fundamental Dirac flavor, saturating the instanton
fermion zero modes with one mass insertion gives a factor
proportional to $m_F(r)$, and hence to $r$, at leading order in
that insertion. This motivates the linear scalar operator only when the dominant
instantons obey $m_F\rho\ll1$. For the heavy-fermion, confining
example discussed in Section~\ref{sec:hidden-dynamics}, that
condition fails at the strong scale. Neither the dilute sum nor its
linear mass-insertion coefficient is then a matching calculation.
In the effective potential used here, $\kappa$, including the Yukawa factor,
remains the effective input specified in
Eq.~\eqref{eq:instanton-coefficient}. The Jacobian determines the
angle; it does not determine the magnitude of $\kappa$.

At $r=v_\phi$ and $\kappa>0$, minimizing the energy with
respect to the dynamical field $\chi$, while keeping $\theta_X$
fixed, selects
\begin{equation}
 \vartheta=0\pmod{2\pi},\qquad
 \chi_0=\sigma_\chi\theta_X\pmod{2\pi\sigma_\chi},\qquad
 \chi_{\rm fl}(x)=\chi(x)-\chi_0.
 \label{eq:alt-vacuum-minimum}
\end{equation}
The subscript $\mathrm{fl}$ denotes the fluctuation about this
minimum. Its kinetic term is unchanged by the constant shift.
After subtracting the minimum energy, the angular potential is
\begin{equation}
 V_{\rm ang}(\chi_{\rm fl})=
 \lambda_{\mathrm I}v_\phi^4
       \left[1-\cos\!\left(\frac{\chi_{\rm fl}}{\sigma_\chi}\right)\right],
 \qquad
 m_\chi^2=\frac{\lambda_{\mathrm I}v_\phi^4}{\sigma_\chi^2}.
 \label{eq:alt-vacuum-cosine}
\end{equation}
The main text chooses the phase origin at this minimum and writes
$\chi$ for $\chi_{\rm fl}$. It therefore uses exactly this
instanton potential and its fixed-$h_\phi$ equation of motion.
The full confining vacuum energy and
its radial derivatives would have to be matched before identifying
this potential with the dynamics of the unbroken hidden gauge theory.

\section{Kinematics of dark matter}
\label{app:dark-matter-kinematics}

This appendix develops the kinematics used in the emission discussion
from the relativistic energy--momentum relation onward. It separates
the selected incoming speed from the speed of the emitted axi-Higgs,
derives both the exact and nonrelativistic production boundaries,
and explains how the illustrative horizontal lines in the main-text
mass plots are related to a Galactic speed distribution. The detailed
derivation is included so that every numerical speed and threshold in
Section~\ref{sec:scattering} can be reconstructed from the stated inputs.

\input{kinematics_appendix}

\end{document}

%% file: kinematics_appendix.tex
\subsection{Production threshold and benchmark kinematics}
\label{kin:sec:setup}

We consider
\begin{equation}
 \Psi(p)+A(P)\longrightarrow\Psi(p')+A(P')+\chi(k),
 \label{kin:eq:reaction}
\end{equation}
where $A$ is an initially stationary xenon nucleus and $\chi$ is the emitted axi-Higgs. The laboratory frame is the initial nuclear rest frame; starred quantities refer to the centre-of-momentum frame. The target is assumed to remain in its ground state, so its mass is the same before and after the collision.

We use natural units in the invariant calculation and write $\beta=v/c$ when converting to conventional speeds, with $c=299792.458~\kms$. The incident speed $v$ belongs to $\Psi$ relative to the target. It does not specify the speed of the emitted $\chi$, which varies over the three-body final-state phase space. For later use, the exact relation between the kinetic energy of a particle and its speed is
\begin{equation}
 T=(\gamma-1)m,\qquad
 \gamma=1+\frac{T}{m},\qquad
 v=c\sqrt{1-\gamma^{-2}}.
 \label{kin:eq:energy-speed}
\end{equation}
The numerical calculation uses one representative xenon nuclear mass; isotope-dependent changes are correspondingly small.

With metric signature $(+---)$, the target-rest-frame invariant is
\begin{equation}
 p^\mu=(E,\mathbf p),\qquad P^\mu=(m_A,\mathbf0),\qquad
 s=(p+P)^2=m_\Psi^2+m_A^2+2m_AE.
 \label{kin:eq:s-lab}
\end{equation}
The minimum final-state energy in the centre-of-momentum frame is the sum of the three final masses. Production therefore requires
\begin{equation}
 \sqrt{s}\geq m_\Psi+m_A+m_\chi.
 \label{kin:eq:exact-threshold-root}
\end{equation}
At equality the final particles have no relative momentum in this frame. Combining the last two equations and writing $T=E-m_\Psi$ gives the exact incident threshold
\begin{equation}
 T_{\min}=m_\chi\left(1+\frac{m_\Psi}{m_A}\right)
                   +\frac{m_\chi^2}{2m_A}.
 \label{kin:eq:Tmin}
\end{equation}
The corresponding incoming speed follows directly from Eq.~\eqref{kin:eq:energy-speed}:
\begin{equation}
 \gamma_{\min}=1+\frac{T_{\min}}{m_\Psi},\qquad
 v_{\min}=c\sqrt{1-\frac{1}{\gamma_{\min}^2}}.
 \label{kin:eq:exact-vmin}
\end{equation}
These equations determine the incident $\Psi$ speed. For a fixed incident speed, the largest producible mass is
\begin{equation}
 m_{\chi,\max}(v)
 =\sqrt{m_\Psi^2+m_A^2+2m_A\gamma(v)m_\Psi}
  -m_\Psi-m_A.
 \label{kin:eq:mmax-exact}
\end{equation}
For $M=m_\Psi+m_A$ this can be written in a numerically stable form,
\begin{equation}
 m_{\chi,\max}=\frac{2m_AT}{\sqrt{M^2+2m_AT}+M}.
 \label{kin:eq:mmax-rationalized}
\end{equation}

For $\beta\ll1$ and $m_\chi\ll m_\Psi,m_A$, Eqs.~\eqref{kin:eq:Tmin} and~\eqref{kin:eq:energy-speed} reduce to
\begin{equation}
 \mu_{\Psi A}=\frac{m_\Psi m_A}{m_\Psi+m_A},\qquad
 m_{\chi,\max}\simeq\frac12\mu_{\Psi A}\beta^2,\qquad
 v_{\min}\simeq c\sqrt{\frac{2m_\chi}{\mu_{\Psi A}}}.
 \label{kin:eq:nonrel-threshold}
\end{equation}
The reduced mass appears because the nonrelativistic laboratory kinetic energy separates into collective and relative motion,
\begin{equation}
 \frac12m_\Psi v^2
 =\frac12(m_\Psi+m_A)V_{\CM}^2
  +\frac12\mu_{\Psi A}v^2.
 \label{kin:eq:energy-split}
\end{equation}
Only the relative part can increase the centre-of-momentum energy above the initial rest masses. For $m_\Psi\gg m_A$, as in the benchmark, $\mu_{\Psi A}\simeq m_A$ and only a fraction $m_A/(m_\Psi+m_A)$ of the laboratory kinetic energy is available for producing additional rest mass.

The main text uses the illustrative masses and incoming speed
\begin{equation}
 m_\Psi=1100~\mathrm{GeV},\qquad m_A=122~\mathrm{GeV},
 \qquad \beta_{\rm ref}=7.7\times10^{-4}.
 \label{kin:eq:inputs}
\end{equation}
These inputs give
\begin{equation}
 \mu_{\Psi A}=109.81997~\mathrm{GeV},\qquad
 v_{\rm ref}=230.8402~\kms,\qquad
 \frac12\mu_{\Psi A}\beta_{\rm ref}^2=32.55613~\mathrm{keV}.
 \label{kin:eq:32kev}
\end{equation}
The exact value from Eq.~\eqref{kin:eq:mmax-rationalized} is $32.55614$ keV. The laboratory kinetic energy is $326.095$ keV, but only the fraction $\mu_{\Psi A}/m_\Psi=0.09984$ contributes to relative motion.

Applying Eqs.~\eqref{kin:eq:Tmin} and~\eqref{kin:eq:exact-vmin} to the two benchmark masses gives
\begin{center}
\begin{tabular}{ccc}
\toprule
$m_\chi$ & $T_{\min}$ & $v_{\min}$\\
\midrule
$6.23$ keV & $62.4021$ keV & $100.98~\kms$\\
$340$ keV & $3405.5742$ keV & $745.99~\kms$\\
\bottomrule
\end{tabular}
\end{center}
Thus the $231~\kms$ reference collision can produce the $6.23$ keV state but not the $340$ keV state.

\begin{figure}[tbp]
\centering
\begin{tikzpicture}
\begin{axis}[width=.91\textwidth,height=7.5cm,xmin=0,xmax=850,ymin=0,ymax=430,
 xlabel={Incoming $\Psi$ speed $v$ [$\kms$]},ylabel={Maximum emitted mass $m_{\chi,\max}$ [keV]},
 grid=major,grid style={gray!18},legend style={font=\small,draw=none,at={(.02,.98)},anchor=north west}]
\addplot[blue!65!black,thick,domain=0:850,samples=140]{0.5*109.8199672668*(x/299792.458)^2*1e6};
\addlegendentry{$\frac12\mu_{\Psi A}(v/c)^2$}
\addplot[black,densely dashed] coordinates {(230.8402,0) (230.8402,430)};
\addlegendentry{Chosen incoming speed}
\addplot[orange!85!black,dashed] coordinates {(0,6.23) (850,6.23)};
\addlegendentry{$m_\chi=6.23$ keV}
\addplot[red!75!black,dashed] coordinates {(0,340) (850,340)};
\addlegendentry{$m_\chi=340$ keV}
\addplot[only marks,mark=*,black] coordinates {(230.8402,32.55613) (100.9809,6.23) (745.9909,340)};
\node[anchor=south west,font=\small] at (axis cs:238,35){$32.6$ keV};
\end{axis}
\end{tikzpicture}
\caption{Calculated production boundary for $m_\Psi=1100$ GeV and $m_A=122$ GeV. At a given speed, masses below the blue curve have open phase space. The two horizontal mass lines intersect the boundary at incoming speeds of approximately $101$ and $746~\kms$. The chosen speed of $231~\kms$ gives the $32.6$ keV limit used in the main-text plots. This calculated diagram is not a halo speed distribution or an experimental exclusion curve.}
\label{kin:fig:boundary}
\end{figure}
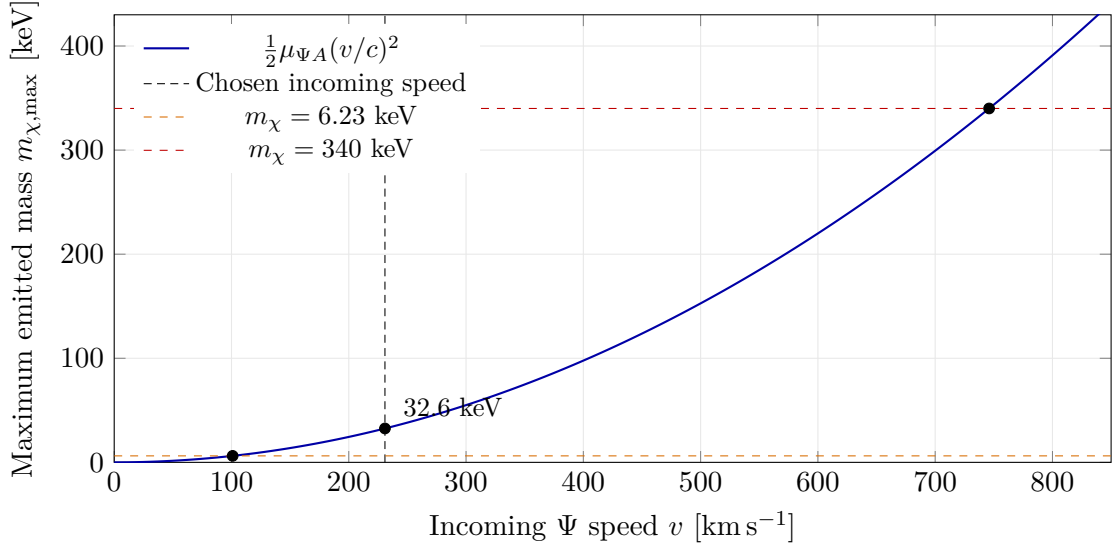

\subsection{The speed of the emitted axi-Higgs}
\label{kin:sec:outgoing}

Fixing the incident energy does not determine the energy of one particle in a three-body final state. For any specified final configuration,
\begin{equation}
 E_\chi^2=m_\chi^2+|\mathbf k|^2,
 \qquad
 \frac{v_\chi}{c}=\frac{|\mathbf k|}{E_\chi}
                =\sqrt{1-\frac{m_\chi^2}{E_\chi^2}},
 \label{kin:eq:outgoing-speed}
\end{equation}
with all quantities evaluated in the same frame. To obtain the kinematic endpoints, combine the recoil particles into
\begin{equation}
 Q=p'+P',\qquad W^2=Q^2.
\end{equation}
where $W\geq M\equiv m_\Psi+m_A$. Since $Q=(\sqrt{s},\mathbf0)-k$ in the centre-of-momentum frame,
\begin{equation}
 E_\chi^*=\frac{s+m_\chi^2-W^2}{2\sqrt{s}},\qquad
 m_\chi\leq E_\chi^*\leq
 E_{\chi,\max}^*=\frac{s+m_\chi^2-M^2}{2\sqrt{s}}.
 \label{kin:eq:Echi-range}
\end{equation}
These are phase-space boundaries; their population depends on the matrix element. At the production threshold $\sqrt{s}=M+m_\chi$, the interval collapses to $E_\chi^*=m_\chi$, so $\chi$ is at rest in the centre-of-momentum frame but not generally in the laboratory.

Choose the incident direction as the $z$ axis. The velocity of the collision centre of momentum relative to the laboratory is
\begin{equation}
 \beta_{\CM}=\frac{|\mathbf p|}{E+m_A},\qquad
 \gamma_{\CM}=\frac{E+m_A}{\sqrt{s}}.
 \label{kin:eq:CM-boost}
\end{equation}
A boost along the incident direction gives
\begin{align}
 E_\chi&=\gamma_{\CM}
     (E_\chi^*+\beta_{\CM}|\mathbf k^*|\cos\theta^*),\\
 k_z&=\gamma_{\CM}
     (|\mathbf k^*|\cos\theta^*+\beta_{\CM}E_\chi^*),\\
 |\mathbf k_\perp|&=|\mathbf k^*|\sin\theta^*.
 \label{kin:eq:chi-boost}
\end{align}
Here $\theta^*$ is measured relative to the boost axis. At threshold all three final particles share the centre-of-momentum velocity; above threshold the emitted-particle speed spans a range.

Above threshold a light emitted particle can be relativistic even when the incoming heavy particle is nonrelativistic. For $m_\chi=6.23$ keV and $v=230.8402~\kms$, Eq.~\eqref{kin:eq:Echi-range} gives $E_{\chi,\max}^*\simeq32.5561$ keV. The upper speed endpoint is therefore
\begin{equation}
 \frac{v_{\chi,\max}^*}{c}
 =\sqrt{1-\left(\frac{6.23}{32.5561}\right)^2}
 \simeq0.982.
\end{equation}
The incoming $\Psi$ is slow because its kinetic energy is small compared with its $1100$ GeV rest energy. The same collision energy is several times the rest energy of a $6.23$ keV axi-Higgs. A low-speed approximation for the incoming particle consequently does not justify setting $E_\chi=m_\chi$ throughout a three-body rate calculation. That substitution is appropriate only at the production boundary.

No outgoing axi-Higgs speed was selected to draw the horizontal lines in the mass plots. Those lines follow from the minimum final energy, allowing all physically permitted final momenta. To predict an outgoing speed distribution one would need the differential matrix element and the three-body phase-space integral.

\subsection{Galactic velocities and the choice of an incoming speed}
\label{kin:sec:halo}

A Galactic interpretation requires a local $\Psi$ population and its velocity distribution, neither of which follows from the particle Lagrangian. The Galactic circular speed sets a characteristic scale for a statistical halo model; individual particles need not follow circular orbits.

Baxter et al.~\cite{Baxter2021} use a truncated Maxwellian halo model and transform it into the laboratory frame. Their reference values include $v_0=238~\kms$, Galactic escape speed $v_{\esc}=544~\kms$, Solar peculiar velocity $(11.1,12.2,7.3)~\kms$, and Earth orbital speed about $29.8~\kms$. These are different quantities: $v_0$ sets a Galactic speed scale; the Solar peculiar velocity corrects the Sun's motion relative to the local circular frame; the Earth's orbital velocity adds a time-dependent laboratory motion.

Let $\mathbf u$ denote a particle velocity in the Galactic rest frame and $\mathbf v$ its laboratory velocity. At these low speeds, the frame transformation is the Galilean vector relation
\begin{equation}
 \mathbf u=\mathbf v+\mathbf V_{\lab}(t),\qquad
 \mathbf V_{\lab}(t)=\mathbf v_0+\mathbf v_{\rm pec}
                      +\mathbf v_{\rm orb}(t).
 \label{kin:eq:halo-boost}
\end{equation}
Relativistic corrections to this frame transformation are negligible. The vector character of Eq.~\eqref{kin:eq:halo-boost} produces the annual modulation; the Earth's orbital speed is not added as a common scalar increment to every particle speed.

Since $|\mathbf u|\leq v_{\esc}$ in a truncated bound-halo model, the triangle inequality gives
\begin{equation}
 |\mathbf v|=|\mathbf u-\mathbf V_{\lab}|
 \leq |\mathbf u|+|\mathbf V_{\lab}|
 \leq v_{\esc}+V_{\lab}.
 \label{kin:eq:lab-endpoint}
\end{equation}
For $V_{\lab}$ of roughly $250~\kms$, this upper envelope is around $800~\kms$. It is a support boundary for this particular halo model, not a speed that most particles have. Consequently a required incident speed of $746~\kms$ can lie within the allowed support while still selecting a sparsely populated tail. The time dependence and halo assumptions affect that tail.

The main text's choice $v_{\rm ref}=230.8402~\kms$ is a representative numerical input on the Galactic speed scale. It is not a value of $v$ deduced from the scalar potential, not the adopted $v_0=238~\kms$, and not a claim about the most probable laboratory speed. Its purpose is to show the emission boundary for one specified collision. Choosing $200$, $300$ or $500~\kms$ would be equally possible as an illustrative calculation, provided the choice were stated and the boundary recomputed.

For the same incident and target masses, the change is quadratic:
\begin{equation}
 m_{\chi,\max}(v)\simeq32.5561~\mathrm{keV}
       \left(\frac{v}{230.8402~\kms}\right)^2.
 \label{kin:eq:rescale}
\end{equation}
The selected speed therefore defines a conditional comparison. A halo prediction instead requires integration over the full speed distribution.

A normalized speed density satisfies
\begin{equation}
 g(v)\geq0,\qquad \int_0^\infty g(v)\,\dd v=1,
 \qquad
 {\cal P}(v_1<v<v_2)=\int_{v_1}^{v_2}g(v)\,\dd v.
 \label{kin:eq:speed-probability}
\end{equation}
The fraction of incident particles above the production threshold is
\begin{equation}
 {\cal P}_{\rm above}(m_\chi)
 =\int_{v_{\min}(m_\chi)}^\infty g(v)\,\dd v.
 \label{kin:eq:above-threshold-fraction}
\end{equation}
For this process $g(v)$ describes the incoming $\Psi$ population. It is not a distribution of outgoing axi-Higgs speeds. The integral measures kinematic accessibility only and does not include the velocity-dependent cross section.

A standard truncated Maxwellian velocity density is
\begin{equation}
 f_{\rm gal}(\mathbf u)=
 \frac{1}{N\pi^{3/2}v_0^3}
 e^{-|\mathbf u|^2/v_0^2}\,
 \Theta(v_{\esc}-|\mathbf u|),\qquad
 N=\operatorname{erf}(z)-\frac{2z}{\sqrt{\pi}}e^{-z^2},\qquad
 z=\frac{v_{\esc}}{v_0}.
 \label{kin:eq:gal-density}
\end{equation}
Here $N$ normalizes the distribution. The associated speed density contains the spherical measure, $g_{\rm gal}(u)=4\pi u^2f_{\rm gal}(u)$. At fixed laboratory speed $V=|\mathbf V_{\lab}|$, the shifted density is $f_{\rm lab}(\mathbf v)=f_{\rm gal}(\mathbf v+\mathbf V_{\lab})$. Its angular integral gives
\begin{equation}
 g_{\lab}(v)=\frac{v}{N\sqrt\pi v_0V}
 \begin{cases}
 e^{-(v-V)^2/v_0^2}-e^{-(v+V)^2/v_0^2},
       &0\leq v<v_{\esc}-V,\\[3pt]
 e^{-(v-V)^2/v_0^2}-e^{-v_{\esc}^2/v_0^2},
       &v_{\esc}-V\leq v<v_{\esc}+V,\\[3pt]
 0,&v\geq v_{\esc}+V.
 \end{cases}
 \label{kin:eq:lab-speed-density}
\end{equation}
The factor $u^2$ explains why a speed density vanishes at zero speed even though the vector density is largest at $\mathbf u=0$. The laboratory distribution has support only up to $v_{\esc}+V$, and the annual change in $V$ produces the usual modulation. This statistical halo model is unrelated to thermal corrections to the axi-Higgs potential.

\subsection{The horizontal lines in the axi-Higgs mass plots}
\label{kin:sec:massplots}

The curves in Figs.~\ref{fig:mass-action} and~\ref{fig:mass-vev} use the mass derived in Section~\ref{sec:mass-plots},
\begin{equation}
 m_\chi=\sqrt{\lambda_{\mathrm I}}v_\phi
    \sqrt{1+\frac{e_B^2v_\phi^2}{M_1^2}}.
 \label{kin:eq:model-mass}
\end{equation}
The horizontal line has a separate origin: it is the kinematic production boundary in Eq.~\eqref{kin:eq:nonrel-threshold}. The common vertical variable $m_\chi$ permits the comparison.

Figure~\ref{fig:mass-action} varies $\lambda_{\mathrm I}$ at fixed $M_1=10$ TeV and $e_B=0.40$, for three different fixed hidden VEVs. At fixed values of the other parameters, Eq.~\eqref{kin:eq:model-mass} is proportional to $\sqrt{\lambda_{\mathrm I}}$. Reducing the coefficient by a factor of one hundred therefore reduces the mass by ten. The logarithmic axes turn this power law into straight lines of slope $1/2$. The dotted line is $m_{\chi,\max}=32.6$ keV for $m_\Psi=1.10$ TeV, $m_A=122$ GeV and incident $v=230.8402~\kms$. Masses below it can be emitted at that speed; masses above it cannot. The marked $6.23$ keV and $0.340$ MeV examples lie on opposite sides.

Figure~\ref{fig:mass-vev} has the hidden Higgs VEV on its horizontal axis, not a speed or an energy supplied to the collision. The potential coefficient is fixed at $\lambda_{\mathrm I}\simeq1.2664\times10^{-14}$ and $e_B=0.40$. The three solid curves use $M_1=1$, $10$ and $100$ TeV.

For $e_Bv_\phi/M_1\ll1$, Eq.~\eqref{kin:eq:model-mass} approaches the dashed curve $m_\chi=\sqrt{\lambda_{\mathrm I}}v_\phi$. In the opposite limit,
\begin{equation}
 m_\chi\simeq\sqrt{\lambda_{\mathrm I}}\frac{e_B}{M_1}v_\phi^2.
\end{equation}
The logarithmic slope changes from approximately one to approximately two. The curve with smaller $M_1$ bends upward earlier because the ratio $e_Bv_\phi/M_1$ becomes large sooner.

The dotted line is horizontal because its calculation holds $m_\Psi$, $m_A$ and $v$ fixed:
\begin{equation}
 m_{\chi,\max}\simeq\frac12
   \frac{m_\Psi m_A}{m_\Psi+m_A}\left(\frac vc\right)^2.
\end{equation}
None of $v_\phi$, $M_1$ or $\lambda_{\mathrm I}$ appears explicitly in this expression. These parameters change the model mass being compared with the boundary; they do not change the energy of the selected incident collision when the masses and speed of the incoming particles are fixed.

In the model, $m_\Psi=y_\Psi v_\phi/\sqrt2$. Holding $m_\Psi=1.10$ TeV fixed while varying $v_\phi$ therefore means adjusting $y_\Psi=\sqrt2m_\Psi/v_\phi$. The horizontal overlay adopts this fixed-incident-mass comparison. If instead $y_\Psi$ were held fixed, then $m_\Psi$ and the reduced mass would change along the scan, and the kinematic boundary would generally cease to be horizontal. The scalar-mass curves and this overlay do not assert perturbativity or phenomenological viability for every associated Yukawa coupling across the full plotting range.

At the marked point Eq.~\eqref{kin:eq:model-mass} gives $m_\chi=0.34003$ MeV. The available $32.6$ keV is too small to produce that rest mass at the plotted reference speed. The required increase in incident speed is approximately
\begin{equation}
 \frac{v_{\min}}{v_{\rm ref}}
 \simeq\sqrt{\frac{340}{32.5561}}\simeq3.23,
\end{equation}
which gives approximately $746~\kms$. Being above the line therefore says that a chosen collision is too slow; it is not a universal upper bound on the existence of that mass.

The dashed mass curve and dotted horizontal line have different origins. The dashed curve approximates a mass formula. The dotted line is a production boundary for one collision setup. Neither is a detector's recoil-energy threshold, a measured exclusion limit, or a chosen outgoing axi-Higgs speed.

\subsection{From allowed kinematics to scattering events}
\label{kin:sec:rates}

A particle with $v>v_{\min}$ is kinematically capable of emission, but the event rate also depends on the amplitude. For local density $n_\Psi$, speed density $g_{\lab}$ and total emission cross section $\sigma_{\rm em}$, the rate per target nucleus is
\begin{equation}
 R_{\rm per\ nucleus}
 =n_\Psi\int_{v_{\min}}^\infty
        g_{\lab}(v)\,v\,\sigma_{\rm em}(v)\,\dd v.
 \label{kin:eq:rate}
\end{equation}
Equation~\eqref{kin:eq:above-threshold-fraction} counts a fraction of incident particles, whereas Eq.~\eqref{kin:eq:rate} weights the distribution by flux and cross section. Detector efficiencies and cuts would enter a differential event-rate calculation.

The main text identifies the emission diagrams but does not supply $\sigma_{\rm em}(v)$. The kinematic comparison therefore determines which incident speeds are capable of emission without assigning an event rate or the final-state speed distribution. A halo interpretation additionally requires an assumed or calculated local $\Psi$ population; the existence of a stable particle in the action does not fix its local density.

A detector recoil threshold is a separate condition. For ordinary nonrelativistic elastic scattering,
\begin{equation}
 E_R=\frac{2\mu_{\Psi A}^2\beta^2}{m_A}\sin^2(\theta^*/2),
 \qquad E_R^{\max}=\frac{2\mu_{\Psi A}^2\beta^2}{m_A}.
\end{equation}
A recoil cut therefore imposes a detector condition distinct from the production condition $m_\chi\leq\mu_{\Psi A}\beta^2/2$. In the three-body process the recoil and emitted energy must be treated together.

The numerical speeds used in the discussion can now be assigned to their respective quantities. The $231~\kms$ value is a selected incoming $\Psi$ speed. The $101$ and $746~\kms$ values are calculated minimum incoming speeds for two different emitted masses. The outgoing axi-Higgs speed varies over final-state phase space and is obtained from its energy and momentum in the specified frame. The horizontal mass-plot line fixes the largest producible rest mass at one incoming speed; a halo calculation integrates over incoming speeds and an emission calculation integrates over outgoing states.